\documentclass[apj]{emulateapj}

\usepackage{natbib}
\usepackage{xspace}

\newcommand{\hii}{{H{\scriptsize II}}\xspace}
\newcommand{\kms}{ km\,s$^{-1}$\xspace}
\newcommand{\miriad}{{\sc Miriad}\xspace}
\newcommand{\um}{{ $\mu{}m$}\xspace}
\newcommand{\degree}{$^{\circ}$\xspace}

\shorttitle{OH masers from the SPLASH survey III}
\shortauthors{Qiao et al.}

\begin{document}

\title{Accurate OH maser positions from the SPLASH survey III: The final 96 square degrees}

\author{Hai-Hua Qiao\altaffilmark{1, 2}, Shari L. Breen\altaffilmark{3, 4}, Jos\'e F. G\'omez\altaffilmark{5}, J. R. Dawson\altaffilmark{4}, Andrew J. Walsh\altaffilmark{4}, James A. Green\altaffilmark{6}, Simon P. Ellingsen\altaffilmark{7}, Hiroshi Imai\altaffilmark{8, 9} and Zhi-Qiang Shen\altaffilmark{2, 10}}
\altaffiltext{1}{National Time Service Center, Chinese Academy of Sciences, Xi'An, Shaanxi, China, 710600; qiaohh@ntsc.ac.cn}
\altaffiltext{2}{Shanghai Astronomical Observatory, Chinese Academy of Sciences, 80 Nandan Road, Shanghai, China, 200030} 
\altaffiltext{3}{Sydney Institute for Astronomy (SIfA), School of Physics, University of Sydney, NSW 2006, Australia}
\altaffiltext{4}{Department of Physics and Astronomy and MQ Research Centre in Astronomy, Astrophysics and Astrophotonics, Macquarie University, NSW 2109, Australia}
\altaffiltext{5}{Instituto de Astrof\'{\i}sica de Andaluc\'{\i}a, CSIC, Glorieta de la Astronom\'{\i}a s/n, E-18008, Granada, Spain} 
\altaffiltext{6}{CSIRO Astronomy and Space Science, Australia Telescope National Facility, PO Box 76, Epping, NSW 2121, Australia}
\altaffiltext{7}{School of Physical Sciences, Private Bag 37, University of Tasmania, Hobart 7001, TAS, Australia}
\altaffiltext{8}{Center for General Education, Kagoshima University, 1-21-30 Korimoto, Kagoshima 890-0065, Japan }
\altaffiltext{9}{Department of Physics and Astronomy, Graduate School of Science and Engineering, Kagoshima University, 1-21-35 Korimoto, Kagoshima 890-0065, Japan}
\altaffiltext{10}{Key Laboratory of Radio Astronomy, Chinese Academy of Sciences, China}

\begin{abstract}
We present high spatial resolution observations of ground-state OH masers achieved with the Australia Telescope Compact Array (ATCA). These observations targeted 253 pointing centres containing OH maser candidates at all four ground-state OH transitions identified in the Southern Parkes Large-Area Survey in Hydroxyl (SPLASH)
across 96 square degrees of the Southern Galactic plane (332\degree$<l<$334\degree and $-$2\degree$<b<+$2\degree, 344\degree$<l<$355\degree and $-$2\degree$<b<+$2\degree, 358\degree$<l<$4\degree and $+$2\degree$<b<+$6\degree, 5\degree$<l<$10\degree and $-$2\degree$<b<+$2\degree). We detect maser emission towards 236 fields and suggest that 7 out of 17 non-detections are due to the slightly lower sensitivity of the ATCA observations, combined with some temporal variability. The superior resolution provided by the ATCA data has allowed us to identify 362 OH maser sites in the 236 target fields. Almost half (160 of 362) of these masers have been detected for the first time. Comparison between these 362 maser sites with information presented in the literature allowed us to categorize 238 sites as evolved star sites (66\%), 63 as star formation (17\%), eight as supernova remnants and 53 unknown maser sites (15\%). We present analysis of the OH masers across the full SPLASH survey range (176 square degrees) and find that the detection rate of 1.7 GHz radio continuum sources (18\%) is lower than that previously found at 8.2 and 9.2 GHz (38\%). We also find that the velocity separations of evolved star sites with symmetric 1612 MHz maser profiles are generally smaller than those with asymmetric profiles. 
\end{abstract}

\section{Introduction}
\label{introduction}
Ground-state hydroxyl (OH) transitions arise from four hyperfine sub-levels within the lowest rotational state $^{2}{\Pi}_ {3/2}$ ($J = 3/2$) and include four frequencies: 1612.231 ($F = 1 \to 2$), 1665.402 ($F = 1 \to 1$), 1667.359 ($F = 2 \to 2$) and 1720.530 MHz ($F = 2 \to 1$). Intense ground-state OH masers were first detected toward Galactic \hii regions \citep{Wee1965,Gun1965} and have since been found to be relatively common in the vicinity of high-mass star formation (HMSF; e.g., \citealt{Are2000}), the circumstellar envelopes of evolved giant and supergiant stars (e.g., Miras, OH/IR stars and planetary nebulae – PNe; e.g., \citealt{Nge1979}, \citealt{Use2012}), and the interaction regions of supernova remnants (SNRs) and their surrounding molecular clouds (\citealt{GR1968}). OH masers associated with HMSF regions are primarily from the main line transitions, i.e., 1665 and 1667 MHz. Evolved star OH masers are predominantly strong in the 1612 MHz transition, which often shows double-horned spectral profiles  (e.g., \citealt{Sea1997}, \citeyear{Seb1997}, \citeyear{Sea2001}). SNRs are only detected in the 1720 MHz OH masers (e.g., \citealt{Fre1996}).  Furthermore, other astrophysical environments, such as comets (\citealt{Gee1998}) and the centers of active galaxies (\citealt{Bae1982}), have also been detected in ground-state OH maser emission.

Most previous searches for ground-state OH maser emission selected target sources based on criteria such as: infrared (IR) point sources with colors indicative of young high-mass star formation regions (e.g., \citealt{Ede2007}); evolved stars associated with water and/or SiO masers (e.g., \citealt{Lee1995}); SNRs (e.g., \citealt{Fre1996}); maser sources with other OH transitions (e.g., \citealt{Cas2004}). 
In addition to these targeted searches, a number of complete surveys have been performed toward regions of the Galactic plane. These surveys were usually restricted to one or two of the ground-state OH transitions (e.g., the 1665 and 1667 MHz main line transitions or 1612 MHz satellite transition) 
and therefore favour either HMSF regions \citep[e.g.][]{Cae1980,CH1983a,CH1983b,CH1987,Cas1998} or evolved star sites \citep[e.g.][]{Sea1997, Seb1997,Sea2001}. Since these previous surveys suffer from biases, further studies should be carried out to comprehensively understand the complete population of ground-state OH masers. 

\begin{figure*}
\centering
\includegraphics[width=1\textwidth]{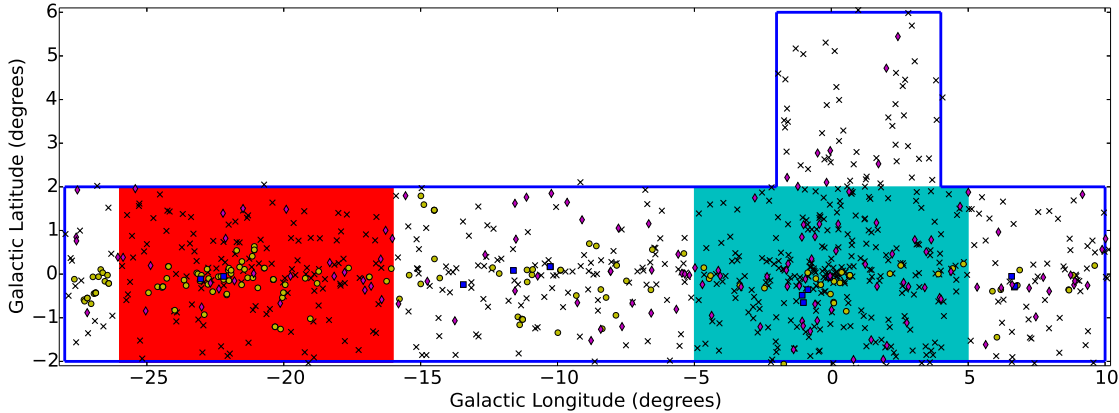}
\caption{The full survey region of SPLASH (blue outline): the red shaded region is the pilot region presented in Paper I while the cyan shaded region is the Galactic Center region which was presented in Paper II. The unshaded area shows the Galactic ranges presented in the current paper. The black crosses, yellow circles, blue squares and carmine rhombuses mark the locations of evolved star, star formation, SNR and unknown OH maser sites we detected in the ATCA observations, respectively.}
\label{glb}
\end{figure*}

The Southern Parkes Large-Area Survey in Hydroxyl (SPLASH) used the Parkes 64 m telescope to completely search a 176 square degree portion of the southern Galactic plane in all four ground-state OH transitions simultaneously (\citealt{Dae2014}). The sensitive OH search covered Galactic longitudes of 332\degree and 10\degree and Galactic latitudes of $-$2\degree and $+$2\degree (152 square degrees), plus an extension around the Galactic Center, i.e., between Galactic longitudes of 358\degree and 4\degree and Galactic latitudes of $+$2\degree and $+$6\degree (24 square degrees)(shown in Figure \ref{glb}). The initial SPLASH single-dish survey produced maser-optimised cubes (see \citealt{Dae2014} for details) with a mean rms (root-mean-square) sensitivity of $\sim$65 mJy (at a velocity resolution of 0.18\kms), allowing us to identify about 600 OH maser sites. The SPLASH observations have allowed us to overcome biases present in a number of previous searches but are limited to a spatial resolution of about 13\arcmin. In order to fully exploit these sensitive OH maser observations, high spatial resolution observations are required, allowing reliable identifications with associated objects. 

We completed the accurate positions of the pilot region (334\degree
$<l<$344\degree and $-$2\degree$<b<+$2\degree, 40 square degrees; \citealt{Qie2016b}, hereafter Paper I; shown in the red shaded region in Figure \ref{glb}) and the Galactic Center region ($-$5\degree$<l<+$5\degree and $-$2\degree$<b<+$2\degree, 40 square degrees; \citealt{Qie2018}, hereafter Paper II; shown in the cyan shaded region in Figure \ref{glb}) with the Australia Telescope Compact Array (ATCA) observations. In these two regions, we detected 571 OH maser sites toward 367 target fields where OH maser candidates were identified in SPLASH. Towards 30 of these target fields, our ATCA observations find no OH maser emission, about half of which might be due to the intrinsic variability of OH masers (Paper I; Paper II). Through comparison with data in the literature, we concluded that, of these 571 OH maser sites, 391 are associated with evolved stars, 95 are associated with star formation, 6 are associated with SNRs and the origins of 79 maser sites remain unknown. Three of the evolved star OH maser sites are associated with PNe, one of which exhibited variable 1720 MHz OH maser emission (IRAS 16333$-$4807; \citealt{Qie2016a}) which may be tracing short-lived equatorial ejections during the PNe formation. This 1720 MHz maser is one of only three such masers ever detected in the vicinity of a PNe (the other two examples are K3-35 -- \citealt{Goe2009} and Vy2-2 -- \citealt{Goe2016}).

This is the third (and final) paper in a series presenting the accurate positions of OH masers identified in the SPLASH survey. The Galactic region presented in the current paper is given in Figure \ref{glb} (96 square degrees; between Galactic longitudes of 332\degree and 334\degree and Galactic latitudes of $-$2\degree and $+$2\degree, between Galactic longitudes of 344\degree and 355\degree and Galactic latitudes of $-$2\degree and $+$2\degree, between Galactic longitudes of 358\degree and 4\degree and Galactic latitudes of $+$2\degree and $+$6\degree, and between Galactic longitudes of 5\degree and 10\degree and Galactic latitudes of $-$2\degree and $+$2\degree). Future work will present the full polarisation characteristics, including an analysis of magnetic fields.

\section{Observations and Data Reduction}
\label{observation}

ATCA observations of 253 fields containing OH maser candidates identified in the SPLASH data were conducted over a series of seven observing sessions during 2015 January 27 -- 28, 2015 January 30 -- 2015 February 1, 2015 March 1 -- 7, 2016 February 19 -- 24, 2016 February 26 -- 29, 2016 March 2 and 2016 March 4. Across these observations the ATCA was in the 6A, 6A, 6C, 6B, 6B, 6B and 6B configurations, respectively, resulting in a synthesised beam in the range 6.5$\arcsec$ $\times$ 4.4$\arcsec$ to 22$\arcsec$ $\times$ 5$\arcsec$. Each of the 253 target fields were observed for a minimum of 20 minutes, conducted as a series of 4-minute integrations distributed across a range of hour angles in order to obtain adequate uv-coverage. 

A full description of the observations and data reduction specifications are provided in Paper I and Paper II, so here we provide only a brief description of the important points. At each epoch, the Compact Array Broadband Backend (CABB) was configured to provide 4 spectral zoom bands (channel spacing of 0.5 kHz or 0.09\kms) but there was some variation in the zoom bandwidths across these epochs. For observations in the 2014 OCT Semester (i.e., 2015 January 27 -- 28, 2015 January 30 -- 2015 February 1 and 2015 March 1 -- 7), we used 2 MHz zoom bands for both the 1612 and 1720 MHz transitions, and 1.5 MHz zoom bands for both the 1665 and 1667 MHz transitions (following our previous observations of the SPLASH pilot region). For observations during the 2015 OCT Semester (i.e., 2016 February 19 -- 24, 2016 February 26 -- 29, 2016 March 2 and 2016 March 4), 4 MHz zoom bands were used for the 1612 MHz transition and 1.5 MHz zoom bands for each of the 1665, 1667 and 1720 MHz transitions. The reduced bandwidths during the Semester 2014 OCT meant that the entire velocity range of six 1612 MHz OH masers (G350.10$+$0.75,  G350.35$−$1.50,  G351.05$+$2.05, G353.65$+$0.80, G353.75$−$1.55 and 353.95$−$1.00) and one 1720 MHz OH maser (G332.275$−$0.05) were excluded in the ATCA observations (see Table \ref{nondetection} for details), and conse-
quently we are not able to provide accurate positions for
these seven sources. Additionally, one 1612 MHz spectrum (G347.396$+$0.394) and three 1667 MHz spectra (G348.668$-$0.715, G351.118$-$0.352 and G008.707$+$0.811) were truncated. At each epoch observations, PKS B1934$-$638 has been used for primary flux density calibration and either PKS B1934$-$638 or PKS B0823$-$500 was used for bandpass calibration. Observations of a nearby phase calibrator (either PKS B1613$-$586, PKS B1710$-$269 or PKS B1740$-$517; chosen to be within 7\degree) were made every $\sim$20 minutes.

The data reduction process followed standard techniques for ATCA spectral line data processing and a detailed description is given in Paper I.  After calibrating the data and producing image cubes, we searched for maser emission in both the cube and peak intensity images. The resultant mean rms of the ATCA observations was 70 mJy in a 0.09\kms channel (c.f., 65 mJy for the Parkes observations in a 0.18\kms channel). The LSR velocity range over which maser emission was searched for in each transition was approximately $-$350 to $+$300\kms ($-$250 to $+$210\kms for the Semester 2014 OCT) for 1612 MHz, $-$160 to $+$180\kms for 1665/7 MHz and $-$120 to $+$200\kms ($-$180 to $+$250\kms for the Semester 2014 OCT) for 1720 MHz. 

Accurate positions of each maser spot were extracted from integrated intensity images (created using the velocity range of the spot) using the \miriad task imfit which also gave a fit error. As discussed in Paper I, we adopted a 1\arcsec\ positional uncertainty for these observations. Spectra at each of the fitted spot positions were extracted from the uv data using the \miriad task uvspec and were used to determine flux density and velocity characteristics. Velocity ranges are defined by emission surpassing 3$\sigma$ in data that was first binned over 5 channels (corresponding to a velocity resolution of about 0.45\kms). In the case of overlapping velocity components with a trough between them, two spectral components are reported if the peak of the weaker feature is stronger than the trough minimum by at least the 1$\sigma$ noise level. As cautioned in Paper I and Paper II, this method may exclude some weaker emission (including line wings) and as such the presented velocity ranges and integrated flux densities should be used as a guide.

We have used the 1720 MHz zoom band (since few masers are detected in this band) to assess the radio continuum properties associated with the maser targets. Data were reduced employing standard techniques for radio continuum data reduction, obtaining a typical rms noise of 10 mJy.

\section{Results}
\subsection{Overall Summary}
\label{results}

We detected OH maser emission towards 236 of the 253 target fields, in total finding 1236 maser spots with peak flux densities between 0.14 and 220 Jy. Section \ref{size} describes the process we undertook to group these maser spots into 362 maser sites according to their separations (see Table \ref{maintable}), 160 of which have been discovered in the SPLASH observations. Among these maser sites, 293 are made up of more than one maser spot and the most extreme source, G351.774$-$0.536, shows a total of 38 maser spots across all four ground-state OH transitions. The majority of the masers sites (298) show emission in the 1612 MHz transition, while 70 show 1665 MHz, 84 show 1667 MHz and 25 show 1720 MHz OH masers, respectively. Section \ref{nodetections} discusses the 17 target fields where we find no maser emission in our ATCA observations.

\begin{figure*}
\includegraphics[width=0.85\textwidth]{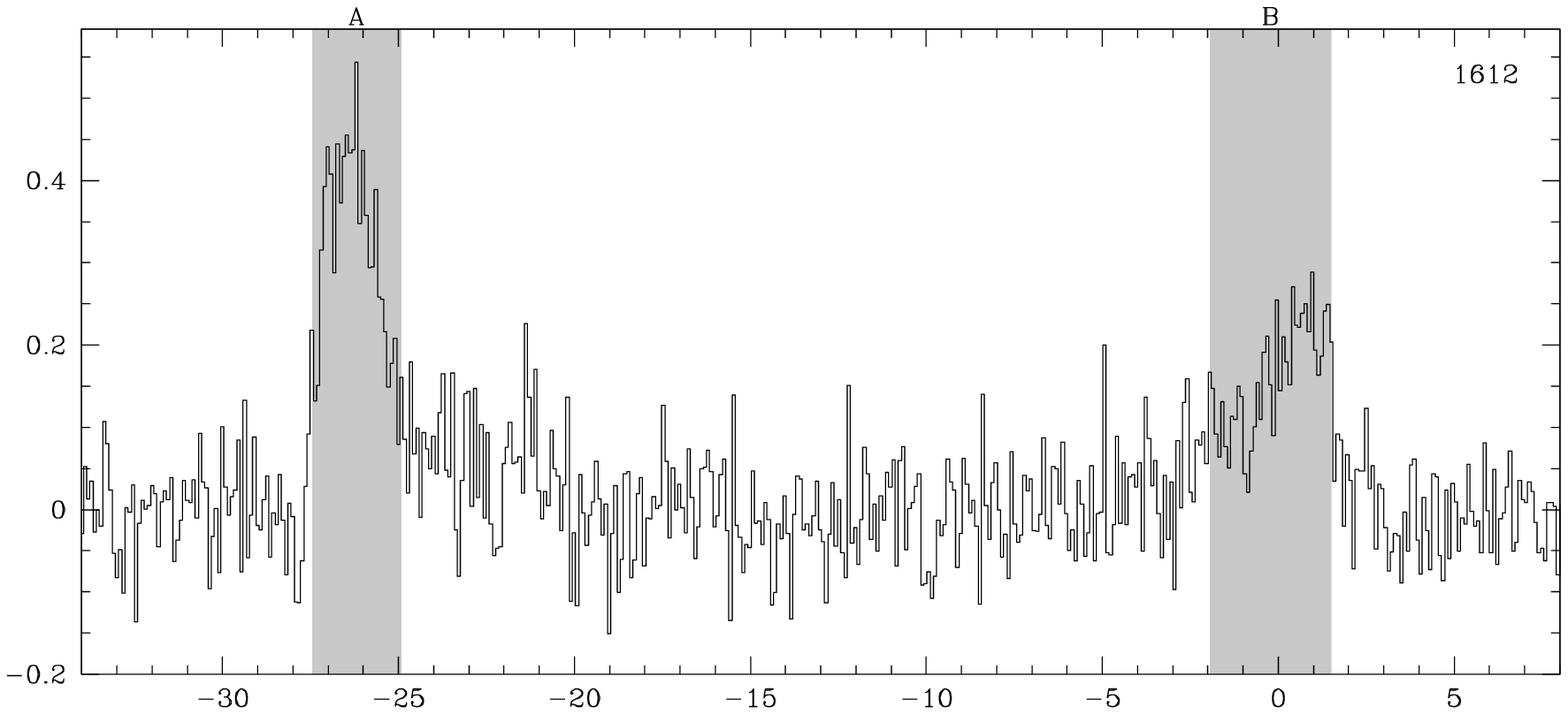}
\includegraphics[width=0.85\textwidth]{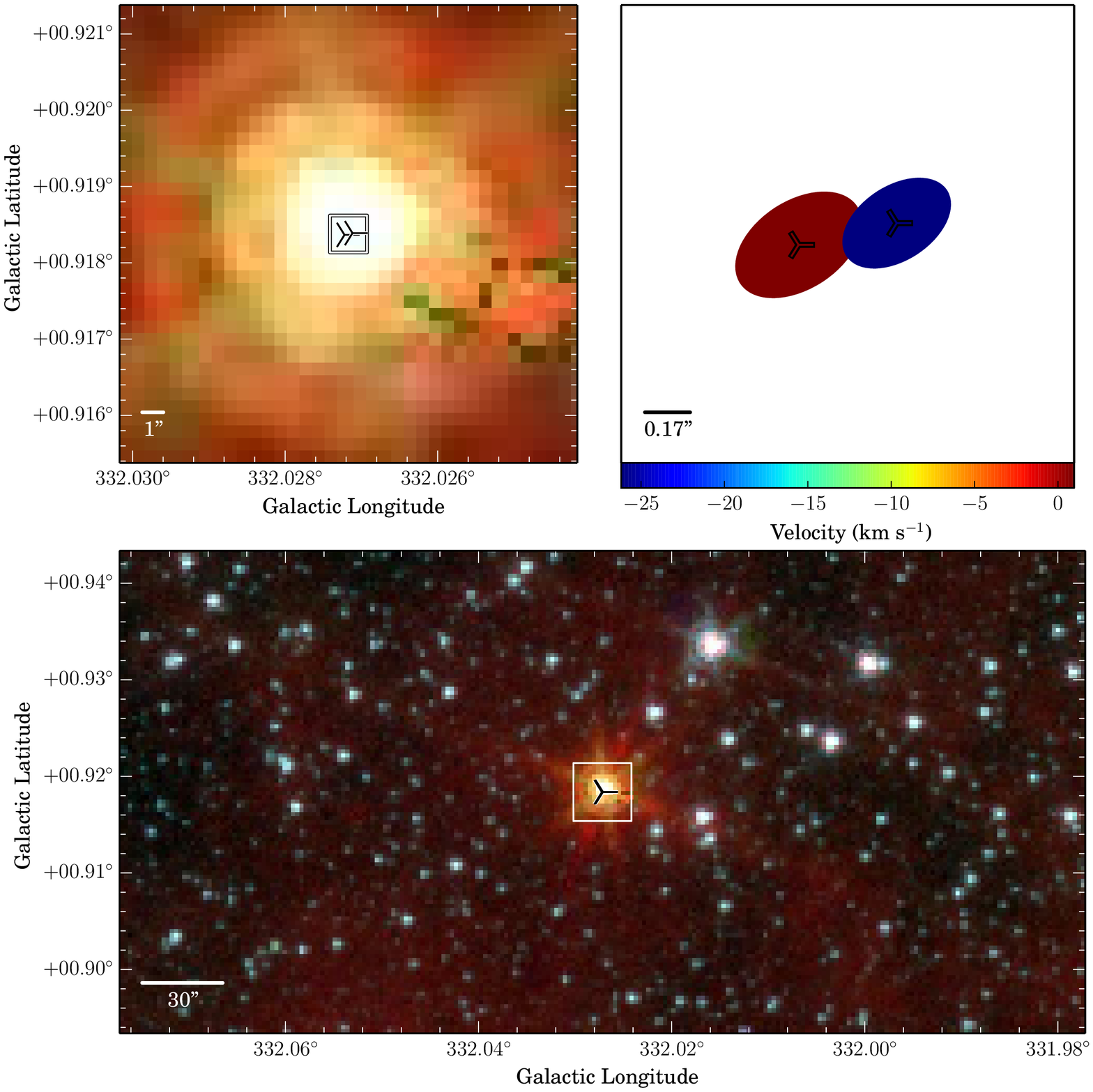}
\caption{G332.027$+$0.918 -- ES. An example figure for one of the detected OH maser sites. Similar figures for each of the 362 detected OH maser sites are available online. In these figures, the upper panel shows the unbinned spectra of the OH maser transitions detected towards each site, with the radial velocity (with respect to the LSR) on the x-axis in units of \kms and flux density (derived from Stokes I) on the y-axis in units of Jy. The shaded regions of the spectrum represent the velocity range over which maser emission is detected. In the bottom, middle-left and middle-right panels of each of the figures, 1612 MHz maser spots are shown with 3-pointed stars, 1665 MHz maser spots with plus symbols, 1667 MHz maser spots with cross symbols and 1720 MHz maser spots with triangles. The bottom panel is a 6$\arcmin$ $\times$ 3$\arcmin$ three-color IR image, usually using GLIMPSE data (blue: 3.6\um; green: 4.5\um; red: 8.0\um) but supplemented by \textit{WISE} (blue: 3.4\um; green: 4.6\um; red: 12.2\um) data for maser sites located beyond the GLIMPSE region. These images are centred on the presented maser site and are marked with all the maser spots detected within the range of the image. The white box shows the extent of the 21.6$\arcsec$ by 21.6$\arcsec$ region shown in the middle-left panel which includes only the spots from the single maser site. The middle-right panel shows the positions and relative error ellipses of all maser spots (derived by the \miriad task \textbf{imfit}) for the maser site, represented by both a colored symbol and a colored error ellipse (coded by velocity of the peak of each maser spot according to the color bar) which represents the relative positional uncertainty of the maser spot.}
\label{G332.027}
\end{figure*}

Following Paper I, we present a figure for each of the maser sites (an example is given in Figure \ref{G332.027}), including spectra of the detected OH maser transitions, a plot of the measured spot positions including a relative error ellipse, and the location of the detected maser spots overlaid on an IR image. Shaded regions of the spectra show the velocity ranges calculated using the method described in Section \ref{observation} and indicate the ranges used to calculate the integrated flux densities given in column 5 in Table \ref{maintable}. In some cases, there is emission in the spectra that is not shaded -- this typically indicates the emission is from nearby unrelated maser sites (details given in Section \ref{individual}) or is a noise spike. Note that three spectra (G005.128$+$1.490, G005.639$+$0.770 and G006.095$-$0.629) have been affected by RFI, which has been flagged, resulting in zero values in some velocity ranges.

\onecolumngrid
\LongTables
\tabletypesize{\tiny}
\tablewidth{\textwidth}
\tablecaption{\textnormal{Details of the 362 OH maser sites, derived from the ATCA observations.}\label{maintable}}
\begin{center}
\begin{deluxetable*}{lccccrrrrrrll}

\tablehead{
\colhead{Name}&\colhead{R.A.}&\colhead{Decl.}&\multicolumn{2}{c}{Flux Density}&\multicolumn{3}{c}{Velocity(\kms)}&\multicolumn{3}{c}{Relative uncertainty}&\colhead{Comments$^{a}$}&\colhead{Ref.$^{b}$}\\
\colhead{}&\colhead{(J2000)}&\colhead{(J2000)}&\colhead{Peak}&\colhead{Integrated}&\colhead{Peak}&\colhead{Min.}&
\colhead{Max.}&\colhead{Minor}&\colhead{Major}&\colhead{Position}&\colhead{}&\colhead{}\\
\colhead{}&\colhead{($^{\rm h\,m\,s}$)}&\colhead{($^{\circ~\prime~\prime\prime}$)}&\colhead{(Jy)}&\colhead{(Jy\kms)}&\colhead{}&\colhead{}
&\colhead{}&\colhead{axis}&\colhead{axis}&\colhead{angle}&\colhead{}&\colhead{}\\
\colhead{}&\colhead{}&\colhead{}&\colhead{}&\colhead{}&\colhead{}&\colhead{}&\colhead{}&\colhead{(arcsec)}&
\colhead{(arcsec)}&\colhead{($^{\circ}$)}&\colhead{}&\colhead{}\\
}
\startdata
G332.027$+$0.918-1612A&16:10:08.860&$-$50:22:42.95&0.54&0.83&$-$26.2&$-$27.5&$-$24.9&0.29&0.46&$-$9.4&ES-VIS&N\\
G332.027$+$0.918-1612B&16:10:08.895&$-$50:22:42.75&0.29&0.57&1.0&$-$2.0&1.5&0.34&0.54&$-$9.4&ES-VIS&N\\
\\
G332.139$-$0.497-1612A&16:16:49.398&$-$51:19:48.39&0.37&0.23&$-$48.5&$-$48.7&$-$47.8&0.45&0.69&$-$11.7&ES-VIS&N\\
G332.139$-$0.497-1612B&16:16:49.355&$-$51:19:47.40&0.49&0.28&$-$18.1&$-$18.4&$-$17.6&0.44&0.67&$-$11.7&ES-VIS&N\\
\\
G332.216$+$0.368-1612A&16:13:22.590&$-$50:39:05.94&0.71&2.04&$-$124.7&$-$129.2&$-$122.8&0.29&0.44&$-$11.8&ES-VIS&N\\
G332.216$+$0.368-1612B&16:13:22.549&$-$50:39:05.90&0.42&1.03&$-$111.5&$-$114.3&$-$110.2&0.30&0.48&$-$11.8&ES-VIS&N\\
\\
G332.265$+$0.309-1612A&16:13:51.502&$-$50:39:39.04&0.33&0.50&$-$74.6&$-$75.1&$-$71.8&0.27&0.40&$-$11.8&ES-VIS&N\\
G332.265$+$0.309-1612B&16:13:51.500&$-$50:39:38.77&0.31&0.24&$-$41.8&$-$41.9&$-$40.5&0.54&0.83&$-$11.8&ES-VIS&N\\
\enddata

\tabletypesize{\footnotesize}
\tablenotetext{}{\textbf{Notes.} The first column lists the name of each maser spot, which is obtained from the Galactic coordinates of the accurate positions, followed by the frequency of the detected OH transition and a letter to denote the sequence of maser spots in the spectrum. The second and third columns are the equatorial coordinates of each maser spot. The fourth and fifth columns list the peak flux density and the integrated flux density. The sixth, seventh and eighth are peak, minimum and maximum velocities, respectively. The ninth, tenth and eleventh columns are minor axis uncertainty, major axis uncertainty and position angle of each maser spot. The twelfth column is the astrophysical identification of each maser site, followed by the reason for the identification (see Section \ref{identification}). The thirteenth column denotes new maser sites with `N' or lists the reference for previously detected maser sites.}

\tablenotetext{a}{Maser sites with unknown associations are listed as U. The remaining sources are described with `Assignment-Reason', where `Assignment' can be SF -- star formation, ES -- evolved star or SN -- supernova remnants. `Reason' can be MMB -- association with a class II methanol maser site which indicates the presence of a high-mass star formation region (\citealt{Bre2013}), based on the Methanol Multibeam survey (\citealt{Cae2010}; \citealt{Gre2010}; \citealt{Cae2011}); HOP -- based on HOPS identification \citep{Wae2014}; RMS -- based on the RMS identifications (\citealt{Lue2013}); VIS -- based on a visual check of GLIMPSE or \textit{WISE} images together with spectral profiles; or based on identifications found through a search of SIMBAD: BER -- \citet{Lee2001}; DAV -- \citet{Dae1993}; DER -- \citet{Dee2002}; KWO -- \citet{Kwe1997}; LIA -- \citet{Lie1989}; LIB -- \citet{Lie1991}; MAT -- \citet{Mae2005}; SAM -- \citet{Sae2009}; SAB -- \citet{Sae2017}; SEA -- \citet{Sea1997}; SEB -- \citet{Seb1997}; SEC -- \citet{Sea2001}; SOS -- \citet{Soe2013}; SUA -- \citet{Sue2006}; ZIJ -- \citet{Zie2001}; CAS -- \citet{Cas1998}; CA4 -- \citet{Cas2004}; COH -- \citet{Coe1995}; GOE -- \citet{Goe2000}; PER -- \citet{Pee1976}; CLA -- \citet{Cle1997}; FRA -- \citet{Fre1996}; KOR -- \citet{Koe1998}; HAS -- \citet{Has1994}.}

\tablenotetext{b}{Maser sites discovered by the SPLASH survey are listed as `N' for new. The remaining maser sites have references to previous observations as follows: BOW (\citealt{BK1989}), CA4 (\citealt{Cas2004}), CAS (\citealt{Cas1998}), CLA (\citealt{Cle1997}), DAV (\citealt{Dae1993}), FRA (\citealt{Fre1996}), KOR (\citealt{Koe1998}), LIA (\citealt{Lie1989}), LIB (\citealt{Lie1991}), LIC (\citealt{Lin1991}), OGB (\citealt{Oge2019}), SEA (\citealt{Sea1997}), SEB (\citealt{Seb1997}), SEC (\citealt{Sea2001}) and ZIJ (\citealt{Zie2001}).}

\tablenotetext{}{(This table is available in its entirety in machine-readable form.)}

\end{deluxetable*}
\end{center}

\twocolumngrid

\subsection{Identification Criteria of Maser Sites}
\label{identification}

SPLASH provides a sensitive census of all four ground-state OH lines allowing us to not only study the association between the lines, but through comparison with complementary data presented in the literature we can also study the OH properties associated with different astrophysical objects. We caution that we cannot account for the possibility that the luminosity distribution of OH masers associated with star formation, evolved star or SNR masers may be different and therefore the sensitivity limit of the SPLASH observations may affect the objects differently.

Section 4.1 of Paper I provides a detailed account of the process undertaken to determine the category that each OH maser belongs to; either evolved stars including PNe, star formation, SNRs or unknown sites. Methanol masers at 6.7 GHz are exclusively associated with HMSF regions (e.g., \citealt{Bre2013}) and a number of our OH maser sites have been identified as star formation sites on the basis of their association with 6.7 GHz masers (\citealt{Cae2010}; \citealt{Gre2010}; \citealt{Cae2011}). Other OH masers have been identified as evolved stars if they show double-horned 1612 MHz emission and are associated with bright star-like emission in GLIMPSE or \textit{WISE}. In the absence of these obvious tracers, other factors are used to determine the association category for each of the OH maser sites, and these are discussed in Section~\ref{individual} and references are also given in column 12 in Table \ref{maintable}.

\subsection{Comments on Individual Sources of Interest}
\label{individual}

\noindent{\textit{G332.418$+$0.848}}. This maser site (associated with IRAS 16084$-$5002) is a new detection with one maser spot at 1612 MHz and one maser spot at 1667 MHz. Inspection of the GLIMPSE three-color image shows that there is a star-like object at the location of the masers. No clear identification in the literature was found for this source; thus, it was identified as an unknown site.

\noindent{\textit{G332.428$+$0.761, G332.821$-$0.883, G345.245$-$0.164, G347.356$+$0.444, G348.447$-$0.384, G348.480$+$0.800, G349.275$-$0.046, G350.724$-$0.655, G351.944$+$0.064, G352.225$+$1.054, G352.729$-$0.666, G353.440$+$0.090, G354.408$+$0.455, G354.648$+$0.001, G359.827$+$2.103, G005.326$-$0.300, G005.639$-$1.615, G006.105$+$0.965, G006.607$-$0.243, G007.472$-$0.979, G007.741$-$0.074, G008.898$-$0.158, G009.360$-$0.260 and G010.018$+$0.810}}. These maser sites show one maser spot at 1612 MHz and are associated with bright star-like objects in the GLIMPSE three-color images, similar to those shown in Figure 9 in Paper I. Two of these sources (G350.724$-$0.655 and G006.105$+$0.965) might be associated with SiO maser emission (\citealt{Dee2000}), but since the SiO observations were conducted with the 45 m Nobeyama radio telescope, interferometric observations would be necessary to confirm the associations. One source (G008.898$-$0.158) is close to a Nova candidate: VVV-NOV-001 (about 1.4\arcsec\ away; \citealt{Sae2012}). \citet{Sae2012} reported a large flux variation in the IR wavelength in this source. However, \citet{Sae2012} cannot rule out this source as a long period variable star or OH/IR star. Therefore, further observations are needed to confirm the nature of this source. As described in Paper I and Paper II, we also used the GLIMPSE IR data to study the properties of any IR point sources located at the maser positions. For those sources with IR point-source counterparts (6/24; G349.275-0.046, G352.729-0.666, G359.827+2.103, G005.639-1.615, G006.105+0.965 and G010.018+0.810), the magnitude of the 4.5\um band is brighter than 7.8 for three sources (out of six sources; the remaining three sources do not have the 4.5\um magnitude), which suggests that these IR point sources are ``obscured'' AGB star candidates and are experiencing very high mass loss (\citealt{Roe2008}). Although it is very possible that these maser sites are associated with the circumstellar envelopes of evolved stars, we do not have any other confirmed identification of their nature. Thus, we include them in the unknown category.

\noindent{\textit{G332.462$+$1.928}}. This maser site only has one maser spot at 1612 MHz and is associated with IRAS 16041$-$4912. In \citet{Has1994}, this source (IRAS 16041$-$4912) was listed as an oxygen-rich AGB star, based on silicate 10\um\ feature. Thus, we classified this source as an evolved star site. In the WISE three-color image, this site is associated with a bright IR star.

\noindent{\textit{G344.448$+$1.797, G348.865$+$1.762, G349.774$+$1.849, G350.896$+$1.246, G351.226$-$1.522, G352.356$-$1.202, G358.386$+$2.214, G001.723$+$2.527, G002.010$+$4.719, G002.433$+$5.443 and G009.173$+$1.826}}. These maser sites only exhibit one maser spot at 1612 MHz. No clear identifications for them were listed in the literature, thus their associations are unknown. In the \textit{WISE} three-color images, they are located in the same position as bright star-like objects. We used the WISE IR data to investigate the properties of any IR point sources located at the same position of these masers. For those sources with WISE IR point-source counterparts (10/11), the magnitude of the 4.6\um band is brighter than 7.8 for ten of them, which indicates these ten IR sources are ``obscured'' AGB star candidates and are experiencing very high mass loss (\citealt{Roe2008}). Although it is very likely that these maser sources are associated with evolved stars, we did not find any other confirmed identification in the literature. Therefore, they are classified as unknown maser sites.

\noindent{\textit{G332.963$-$0.679 and G333.068$-$0.447}}. Emission in the velocity ranges of $-$52 to $-$50\kms (G332.963$-$0.679) and $-$54.5 to $-$51\kms (G333.068$-$0.447) in the 1665 MHz spectra is from the nearby strong maser site G333.135$-$0.432, whose peak flux density is about 43 Jy at $-$51\kms. 

\noindent{\textit{G333.103-1.074}}. The GLIMPSE three-color image in the bottom panel is incomplete due to the observations of the Spitzer Space Telescope.

\noindent{\textit{G333.130$-$0.424}}. The absorption features in the 1665 and 1667 MHz spectra are side-lobe contamination from the nearby source G333.135$-$0.432.
 
\noindent{\textit{G333.466$-$0.164}}. The emission in the velocity range of $-$49 to $-$46\kms in the 1665 and 1667 MHz spectra is from the nearby strong maser site G333.608$-$0.215.

\noindent{\textit{G333.603$-$0.212}}. We detected two 1665 MHz maser spots with peaks at $-$50.3 and $-$48.1\kms at the location of IRAS 16183$-$4958. The unshaded emission in the 1665 MHz spectrum is from the nearby strong star formation maser site G333.608$-$0.215. Previous single dish OH observations targeting IRAS 16183$-$4958 have revealed a weak 4765 MHz OH maser at $-$49.7\kms \citep{Coe1995}. Efforts to derive the precise position of the 4765 MHz OH maser by \citet{DE2002} failed (likely due to intrinsic variability) and so it is unclear if the detected 4765 MHz OH masers is at the location of this new 1665 MHz OH maser or possibly at nearby G333.608$-$0.215.

\noindent{\textit{G344.890$+$0.120}}. This maser site is identified as an evolved star site by \citet{WO1978} and \citet{LC1989} (source WOS69). It is a new detection and only shows one maser spot at 1667 MHz.

\noindent{\textit{G344.905$+$0.115}}. The emission in the velocity range of $-$24 to $-$15\kms is from the nearby strong maser site G344.929$+$0.014, whose flux density is about 117 Jy at $-$23\kms.

\noindent{\textit{G345.003$-$0.225}}. The emission in the velocity range of $-$24 to $-$21.5\kms at 1612 MHz is from the nearby strong maser site G344.929$+$0.014. This star formation maser site shows all four ground-state OH transitions, which was also detected with main line and 1720 MHz transitions by \citet{Cas1998} and \citet{Cas1999}, respectively. This site is also associated with a 6.7 GHz methanol maser (\citealt{Cae2010}).

\noindent{\textit{G345.052$-$1.855}}. This maser source (associated with IRAS 17088$-$4221) is identified as a post-AGB star maser site (\citealt{Sze2007}) and shows 19 maser spots in the 1612, 1665 and 1667 MHz spectra distributed in the velocity range of $-$14 to $+$20\kms. This maser site was also detected by \citet{Seb1997} at 1612 MHz with a double-horned spectral profile. \citet{Dee2007} detected a 22 GHz water maser in the velocity range of $-$5 to $+$16\kms toward this source with the Tidbinbilla 70 m radio telescope. In the GLIMPSE three-color image, this star shows strong emission at 4.5\um.

\noindent{\textit{G345.332$-$0.270}}. The emission in the velocity range of $+$12 to $+$16\kms at 1612 MHz is from the nearby strong source G344.929$+$0.014.

\noindent{\textit{G345.494$+$1.469}}. The emission in the velocity range of $-$15 to $-$14\kms at 1665 MHz is from the nearby strong source G345.498$+$1.467.

\noindent{\textit{G345.495$+$1.461}}. The emission in the velocity range of $-$21 to $-$17\kms at 1720 MHz is from the nearby strong source G345.497$+$1.461.

\noindent{\textit{G345.498$+$1.467}}. The emission in the velocity range of $-$21 to $-$20.5\kms and $-$13 to $-$12\kms at 1665 MHz is from the nearby strong source G345.494$+$1.469.

\noindent{\textit{G345.653$-$1.005}}. This maser site has also been detected by \citet{Seb1997} with a 1612 MHz double-horned profile peaked at $-$13.2 and $-$1.5\kms. The spectral profile at 1612 MHz in our observations deviates from the typical double-horned profile with five maser spots distributed in the velocity range of $-$19.5 to $+$0.5\kms. In the GLIMPSE three-color image, this star is very bright at 8\um.

\noindent{\textit{G345.680$-$1.420}}. The \textit{WISE} three-color image in the bottom panel is incomplete due to the coverage of the \textit{WISE} observations.

\noindent{\textit{G345.928$+$0.175}}. This new maser site is associated with the Mira-type variable star V605 Sco (\citealt{Sae2017}). The 1612 MHz OH spectrum shows two maser spots distributed in the velocity range of $-$71 to $-$62\kms. The 1665 MHz OH spectrum has two maser spots in the velocity range of $-$70 to $-$64\kms. The 1667 MHz OH spectrum exhibits two maser spots in the velocity range of $-$70 to $-$66\kms.

\noindent{\textit{G346.250$-$1.068}}. This new maser site shows two maser spots at 1612 MHz. In the \textit{WISE} three-color image, it appears to be associated with a star-like object. Although it is likely to be an evolved star, we did not find any identification in the literature. Thus it is categorized as an unknown maser site.

\noindent{\textit{G346.560$-$0.239}}. This previously detected 1720 MHz OH maser site is associated with the SNR G346.6$-$0.2 (\citealt{Koe1998}). In the GLIMPSE three-color image, this site is located in the diffuse emission background.  

\begin{figure*}
\includegraphics[width=0.9\textwidth]{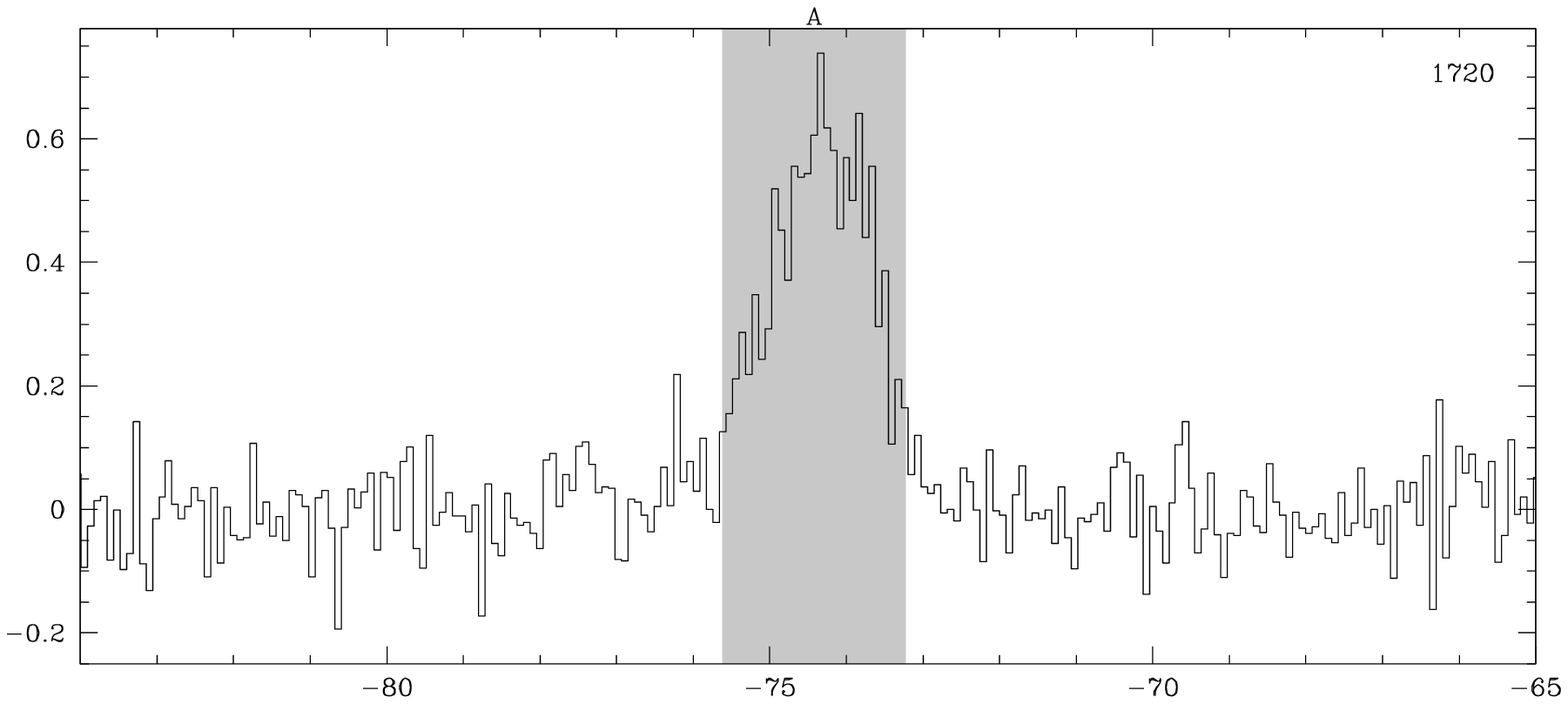}
\includegraphics[width=0.9\textwidth]{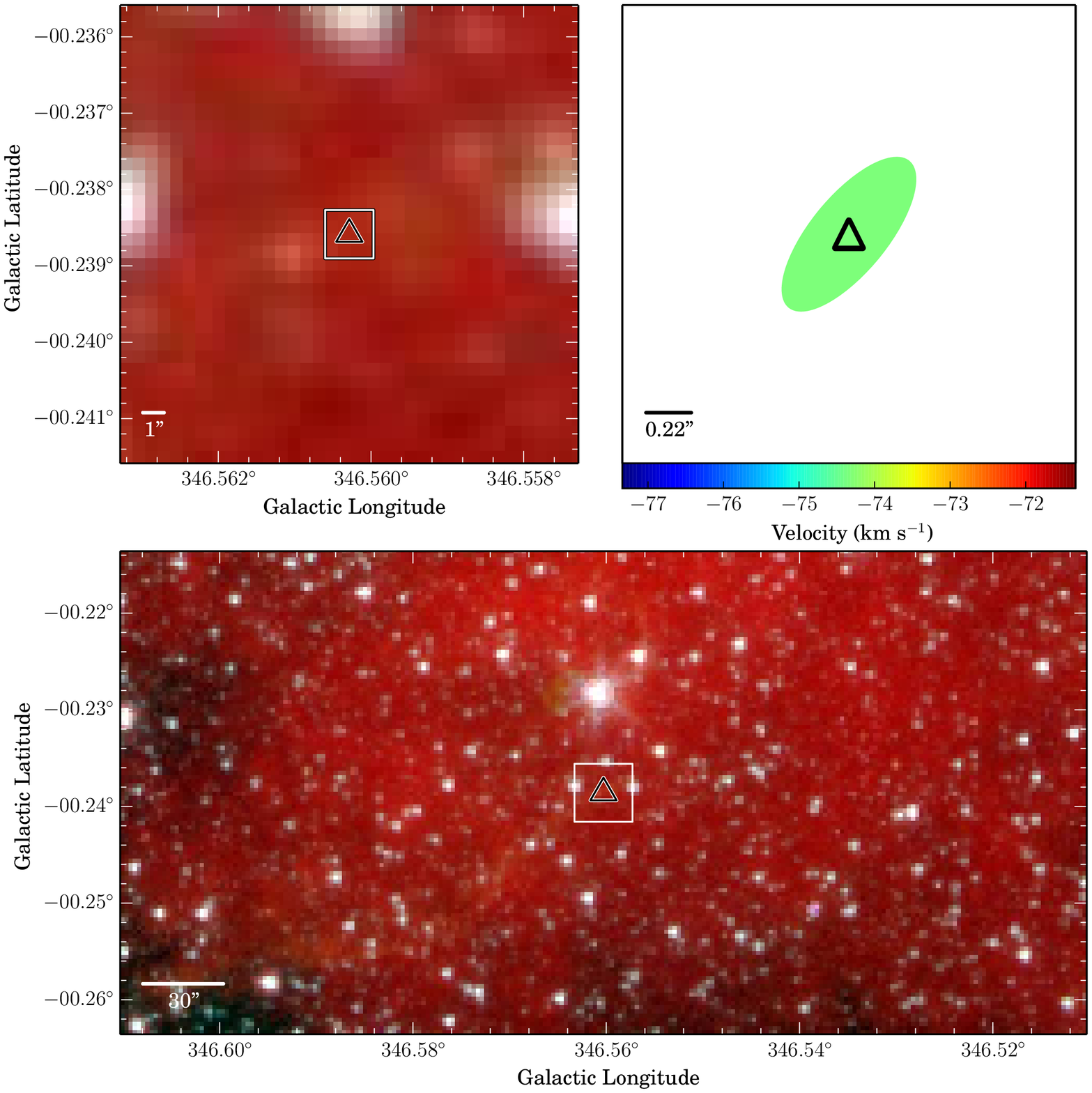}
\caption{G346.560$-$0.239 -- SN}
\label{G346.560}
\end{figure*}

\noindent{\textit{G347.026$-$1.382}}. This maser site (associated with IRAS 17130$-$4029) is identified as an evolved star site based on RMS (\citealt{Lue2013}). In the \textit{WISE} three-color image, the star is very bright at 12\um.

\noindent{\textit{G347.396$+$0.394}}. This maser site was detected and identified as an evolved star maser site by \citet{Seb1997}. The absence of the 1612 MHz spectrum at velocities lower than $-$198\kms is due to the setup of the zoom bands described in Section \ref{observation}. In \cite{Seb1997}, they detected the 1612 MHz double-horned profile peaked at $-$201.6 and $-$176.7\kms.

\noindent{\textit{G347.662$+$1.039}}. This maser site (associated with IRAS 17048$-$3832) was detected as a single-peaked 1612 MHz OH maser by \citet{Seb1997}. In our observations, we detected a double-horned profile at 1612 MHz and two maser spots at 1665 MHz. We identify this source as an evolved star site based on \citet{Kwe1997}.

\noindent{\textit{G348.391$+$0.083}}. This 1720 MHz OH maser corresponds to the brightest maser spot detected towards SNR CTB37A in \citet{Fre1996}, thus is identified as a SNR maser site. The remaining 1720 MHz OH maser spots  associated with SNR CTB37A were either close to or below our sensitivity when detected \citep[flux densities ranging between $\sim$0.1 and $\sim$0.5 Jy,][]{Fre1996}, thus we did not detect them in our observations. In the GLIMPSE three-color image, this source is located within diffuse emission. 

\noindent{\textit{G348.445$+$1.622}}. This maser site shows two maser spots at 1612 MHz. In the \textit{WISE} three-color image, this site is associated with a star-like object. No identification is found in the literature, thus this source is categorized as unknown.

\noindent{\textit{G348.668$-$0.715}}. This maser site has been previously identified to be associated with an evolved star in \citet{Seb1997}. In addition to the double-horned profile at 1612 MHz, we also detected a maser spot at 1667 MHz in the redshifted velocity range. The absence of the 1667 MHz spectrum at velocities lower than $-$168\kms is due to the setup of the zoom bands described in Section \ref{observation}. We checked the 1667 MHz spectrum in the Parkes observation and only found one 1667 MHz maser feature in the redshifted velocity range (peaked at $\sim-$162\kms). Therefore, our ATCA observations included all of the 1667 MHz emission detected with Parkes. Note that this maser site is the source d197 in \citet{Dea2004}. They marginally detected the redshifted component at 1667 MHz, with nothing in the blueshifted velocity range. Further, \citet{Dea2004} also detected a peak at 1665 MHz in the blueshifted velocity range (with a peak velocity of $-$194.08\kms), which is not covered by our observations. In the GLIMPSE three-color image, this star is very bright at 8\um.

\noindent{\textit{G348.698$-$1.028}}. The emission in the velocity range of $-$22 to $-$19\kms and $-$14 to $-$13\kms at 1665 MHz is from the nearby strong source G348.550$-$0.979.

\noindent{\textit{G348.727$-$1.039}}. This 1720 MHz OH maser was also recently detected by Ogbodo et al. (submitted). It is located close to the star formation maser site G348.727$-$1.037, which shows both 1665 MHz OH  (\citealt{Cas1998}) and the 6.7 GHz methanol masers (\citealt{Cae2010}). In the GLIMPSE three-color image, this maser is also associated with an extended green object (EGO; \citealt{Cye2008}), thus is categorized as a star formation source.

\noindent{\textit{G349.180$+$0.203}}. This maser site shows the typical double-horned profile at 1612 MHz, is identified as an evolved star site by \citet{Seb1997} and a long-period variable star V1013 Sco by \citet{Sae2009}. This star also hosts 43 and 86 GHz SiO maser emission (\citealt{Ste2018}).

\noindent{\textit{G349.184$-$1.243}}. This maser site has been detected with a single-peaked spectrum at 1612 MHz by \citet{Seb1997}. \citet{Dea2004} detected a double-horned profile at 1612 MHz (source d200), which is similar to our result. In addition, we also detected two maser spots at 1665 MHz distributed in the blueshifted velocity range. Comparison with RMS (\citealt{Lue2013}) allowed us to identify this source as an evolved star maser site.


\noindent{\textit{G349.273$-$1.216}}. The apparent absorption features in the 1612 MHz spectrum is side-lobe contamination from the nearby source G349.184$-$1.243.

\begin{figure*}
\includegraphics[width=0.9\textwidth]{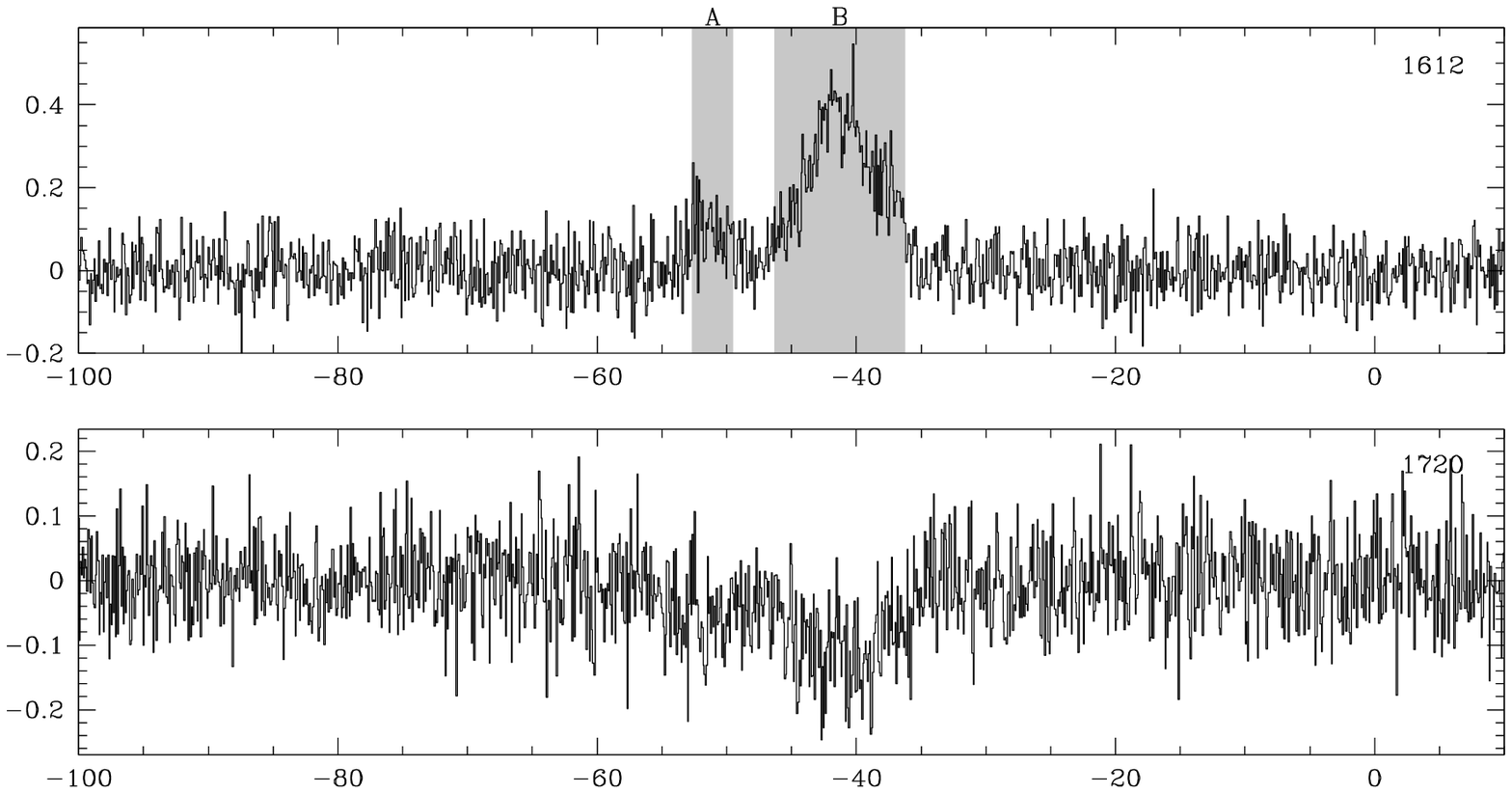}
\includegraphics[width=0.9\textwidth]{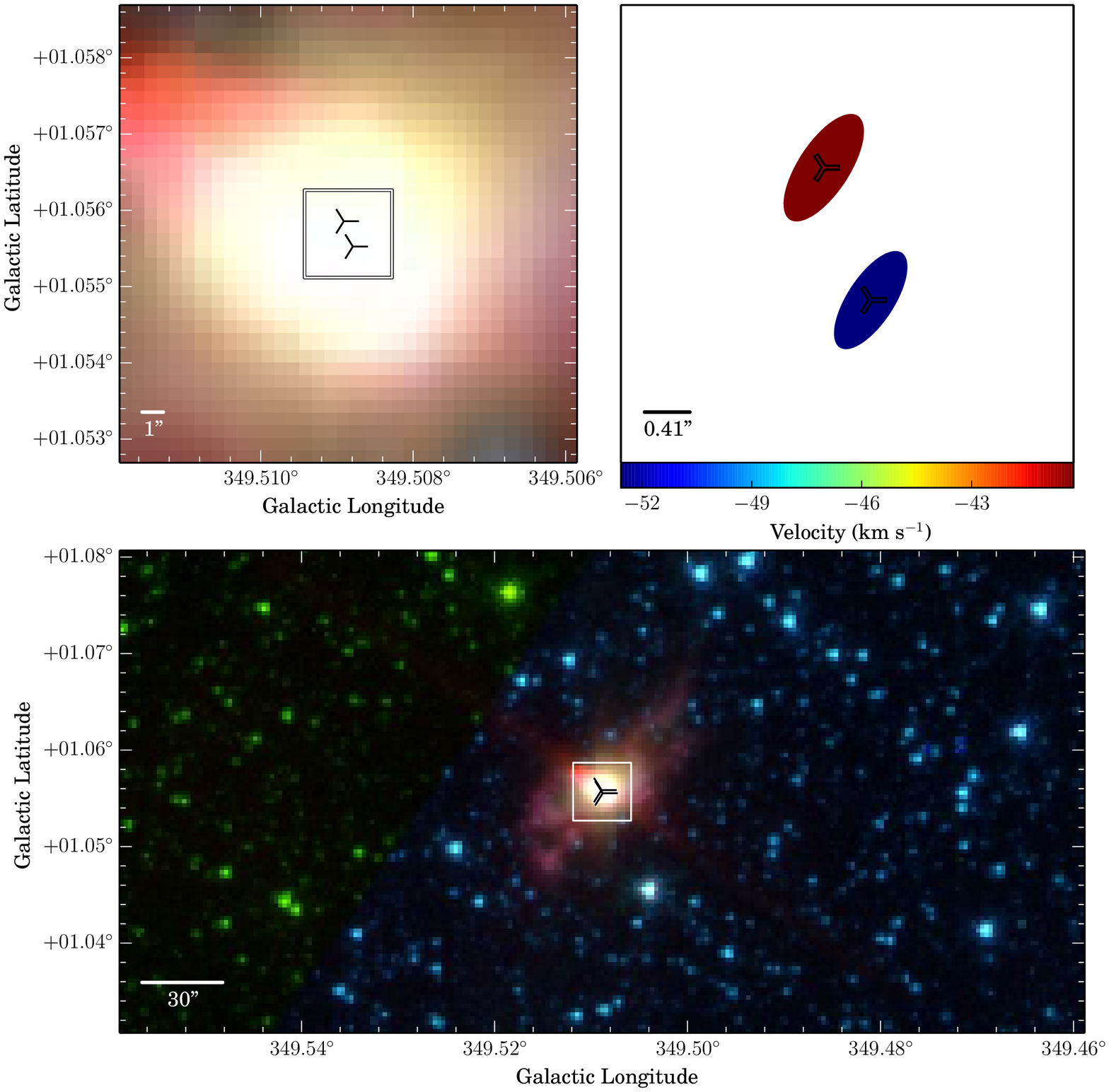}
\caption{G349.509$+$1.055 -- PN -- possible thermal emission}
\label{G349.509}
\end{figure*}

\noindent{\textit{G349.509$+$1.055}}. This OH emission is associated with NGC 6302, a well-known PN. Our OH spectrum at 1612 MHz shows two components, similar to the spectra obtained by other authors (e.g., \citealt{Pae1988} and \citealt{Zie1989}). \citet{Goe2016} suggested that the 1612 MHz OH emission might not be of maser nature, but thermal, given the similarity of the spectrum at 1612 MHz with those of OH at 1720 MHz (but in absorption, as shown in Figure \ref{G349.509}) and CO lines (e.g., \citealt{Pee2007}). This is also suggested by our data, since the OH emission at 1612 MHz is not seen at the longest baselines (those with the antenna 6 of the ATCA). This indicates that the emission is resolved for baselines $\simeq 6$ km, which is not expected for maser components. In the GLIMPSE three-color image, this PN shows bipolar structures.

\noindent{\textit{G349.730$+$0.167, G349.731$+$0.172 and G349.734$+$0.172}}. These three 1720 MHz OH masers are associated with the SNR G349.7+0.2 (\citealt{Fre1996}). For G349.730$+$0.167, the unshaded emission in the velocity range of $+$14.5 to $+$15\kms is from the source G349.734$+$0.172 and the unshaded emission in the velocity range of $+$15.5 to $+$16\kms is from the source G349.731$+$0.172. For G349.731$+$0.172, the unshaded emission at velocities lower than $+$15\kms is from the source G349.734$+$0.172. In the GLIMPSE three-color images, these three maser sites are located in the diffuse emission background.

\noindent{\textit{G349.804$-$0.321 and G350.546$+$0.061}}. These two maser sites were first discovered by \citet{Cae1981}. \citet{Sea1997} detected them with typical double-horned profiles at 1612 MHz and identified them as evolved star sites. In addition to the double-horned profiles at 1612 MHz, we also detected double-horned profiles at 1667 MHz with stronger peak flux densities in the redshifted velocity range, which is opposite to the 1612 MHz OH masers. \citet{Use2012} listed G349.804$-$0.321 as a candidate OHPN (also reported as such by \citealt{GHe2007}), due to the presence of radio continuum emission. In the GLIMPSE three-color images, these stars are very bright at 8\um.

\noindent{\textit{G349.949$+$1.537}}. This maser site was detected with the typical double-horned profile at 1612 MHz by \citet{Sea1997} and was identified as an evolved star site. In addition to the double-horned profile at 1612 MHz, we also detected the double-horned profile at 1667 MHz with stronger peak flux density in the blueshifted velocity range, which is opposite to the 1612 MHz OH maser. In the \textit{WISE} three-color image, this star is very bright at 12\um.

\noindent{\textit{G349.963$-$0.025}}. This maser site has been observed with a single-peaked 1612 MHz spectrum by \citet{Sea1997}. We obtained three maser spots at 1612 MHz. This site is associated with a long-period variable star (V1015 Sco; \citealt{Sae2009}), thus is identified as an evolved star site.

\noindent{\textit{G350.287$+$0.058}}. This site is identified as an evolved star site by \citet{Sea1997}. The apparent absorption features in the 1612 MHz spectrum are side-lobe contamination from the nearby source G349.963$-$0.025.

\noindent{\textit{G350.351$+$1.645}}. This maser site is a new detection with two maser spots at 1612 MHz and is associated with IRAS 17104$-$3601. This source (IRAS 17104$-$3601) has a 9.7\um\ silicate feature in \citet{Kwe1997}, which suggests it is an oxygen-rich evolved star. Thus, we include this source in the evolved star category.

\noindent{\textit{G350.971$+$0.437}}. This maser site was categorized as an evolved star site by \citet{Lie1989}. The apparent absorption features in the 1612 MHz spectrum are due to side-lobe contamination from the nearby source G350.848$+$0.184.

\noindent{\textit{G351.118$-$0.352}}. This known maser site is associated with an evolved star (\citealt{Sea1997}). Its IR colors and variability (\citealt{JE2006}) suggest it is an AGB star. We re-detected the double-horned profile at 1612 MHz (also present in the \citealt{Sea1997} observations) and also found two maser spots at 1667 MHz in the redshifted velocity range. In the Parkes observations, the 1667 MHz OH maser also showed a double-horned profile with spectral peaks at $\sim-$162 and $\sim-$127\kms, similar to the 1612 MHz OH maser. The absence of the 1667 MHz spectrum at velocities lower than $-$160\kms is due to the setup of the zoom bands described in Section \ref{observation}, therefore missing the 1667 MHz maser spot at $\sim-$162\kms.

\noindent{\textit{G351.160$+$0.698}}. The negative features in the velocity range of $-$9.3 to $-$8.8\kms are side-lobe contamination from the nearby source G351.417$+$0.645.

\noindent{\textit{G351.417$+$0.645}}. This source is the strongest OH maser detected in the current Galactic range: 220 Jy at a velocity of $-$8.8\kms at 1665 MHz. This maser site has also been detected by \citet{Cas1998} and is associated with a 6.7 GHz methanol maser (\citealt{Cae2010}), identifying it as a star formation maser site. The emission in the velocity range of $-$15.5 to $-$12.3\kms at 1667 MHz is from the nearby strong source G351.160$+$0.698.

\noindent{\textit{G351.580$-$1.265}}. This 1667 MHz OH maser is a new detection. We did not find any identification in the literature, thus this source is categorized as an unknown maser site. In the \textit{WISE} three-color image, this site is associated with a star-like object.


\noindent{\textit{G351.634$-$1.252}}. This 1612 MHz OH maser was detected by \citet{Lin1991} and is associated with the well-known \hii region G351.63$-$1.25 (\citealt{Pee1976}). Thus, we categorized this source in the star formation category. In the \textit{WISE} three-color image, this source is located in a region of extended emission, which may be the \hii region.

\noindent{\textit{G351.774$-$0.536}}. This star formation OH maser site shows all four ground-state OH transitions and is the richest maser site in this paper (38 maser spots). This source is associated with a 6.7 GHz methanol maser (\citealt{Cae2010}) and was also detected by \citet{Cas1998}. In the GLIMPSE three-color image, this site is associated with several bright EGOs.

\noindent{\textit{G351.932$+$0.064}}. This OH maser is a new detection with two maser spots at 1612 MHz. We did not find any identification in the literature. In the GLIMPSE three-color image, this source is associated with a star-like object. Thus, we identified this source as an unknown maser site.

\noindent{\textit{G352.044$+$0.530}}. This maser site (associated with IRAS 17197$-$3517) is an evolved star site (\citealt{Sea1997}) with a double-horned profile at 1612 MHz and one maser spot in the redshifted velocity range at 1667 MHz. \citet{Dee2004} detected a 43 GHz SiO maser toward this source with the Nobeyama radio telescope.

\noindent{\textit{G352.919$+$0.063}}. This maser site is a new detection with five maser spots at 1665 MHz and one maser spot at 1667 MHz. \citet{Wae2014} detected a water maser toward this source, and considered it is arising from an evolved star, given the absence of dense molecular gas associated with it. Thus, we classified it as an evolved star. In the GLIMPSE three-color image, this site is associated with a bright IR star. It is probably associated with IRAS 17240$-$3449, classified as an O-rich AGB star by \citet{Lee2003}.

\noindent{\textit{G353.252$-$0.243}}. This maser site was detected with two maser spots at 1612 MHz by \citet{BK1989}. We only detected one maser spot at 1612 MHz. We identified this source as an evolved star site based on the classification of the associated water maser in \citet{Wae2014}. In the GLIMPSE three-color image, this source is associated with a bright IR star.

\noindent{\textit{G353.331$+$1.155}}. This maser site is a new detection with only one maser spot at 1612 MHz. We did not find any identification in the literature. In the GLIMPSE three-color image, this source is located in the diffuse emission background. Thus, we included this source in the unknown category.

\noindent{\textit{G353.421$-$0.894}}. This maser site was detected with an irregular (i.e., neither double-horned or single-peaked) 1612 MHz spectrum by \citet{Sea1997}. We detected an irregular spectrum at 1612 MHz and two broad maser spots (line widths $\sim$12\kms) at 1665 MHz. In the GLIMPSE three-color image, this site is associated with a star-like object with bright 8\um\ emission. We did not find any classification of its nature in the literature, thus this source is an unknown maser site.

\noindent{\textit{G353.810$+$1.452}}. This maser site has been detected with a single-peaked spectrum at 1612 MHz by \citet{Sea1997}. We obtained a double-horned profile at 1612 MHz, typical of an evolved star. Combining the \textit{WISE} three-color image (an IR star-like object), we identified this site as an evolved star site.


\noindent{\textit{G354.000$-$0.593}}. This maser site is a new detection with a double-horned profile at 1612 MHz. In the GLIMPSE three-color image, it is associated with a star-like object, which is very bright at 8\um. Based on the 1612 MHz spectrum and the IR image, we included this source in the evolved star category.

\noindent{\textit{G354.104$-$1.982}}. This maser site was detected with a double-horned profile at 1612 MHz by \citet{Sea1997}, who identified the exciting source as an evolved star. It is probably associated with IRAS 17355$-$3457, identified as an O-rich AGB star based on the presence of a 9.7\um silicate dust feature (\citealt{Kwe1997}). We only detected the blueshifted 1612 MHz maser spot. The redshifted 1612 MHz maser spot in \citet{Sea1997} was $\sim$0.27 Jy, which is below our detection limit ($\sim$0.4 Jy). In the \textit{WISE} three-color image, this source is associated with a bright IR star.

\noindent{\textit{G354.340$-$0.806}}. This OH maser is a new detection with two maser spots at 1612 MHz. In the GLIMPSE three-color image, it is associated with a star-like object. We did not find any identification in the literature, thus this source is classified as an unknown maser site.

\noindent{\textit{G354.508$-$1.377}}. This maser site was detected with a double-horned profile at 1612 MHz and was identified as an evolved star site (\citealt{Sea1997}). As well as the double-horned profile at 1612 MHz, we also detected a double-horned profile at 1667 MHz. In the \textit{WISE} three-color image, this star is very bright at 12\um.

\noindent{\textit{G354.545$+$0.020}}. This 1667 MHz OH maser shows one maser spot and is a new detection. In the GLIMPSE three-color image, this site is associated with a star-like object. We did not find any identification in the literature, thus this source is included in the unknown category.

\noindent{\textit{G354.642$+$0.830}}. This OH maser site was detected with a double-horned profile at 1612 MHz and was identified as an evolved star site (\citealt{Sea1997}), with IR colors consistent with a post-AGB object (\citealt{Sev2002}). We only detected the blueshifted component of the double-horned profile at 1612 MHz. The redshifted component detected by \citet{Sea1997} was $\sim$0.2 Jy, which is below our detection limit ($\sim$0.4 Jy). In the GLIMPSE three-color image, this star is very bright at 8\um.

\noindent{\textit{G354.817$-$0.024}}. This new 1612 MHz OH maser shows a double-horned profile in the spectrum. In the GLIMPSE three-color image, this source is not associated with any star-like object. We did not find any identification in the literature. Thus, this source is classified as unknown maser site.

\noindent{\textit{G354.884$-$0.539}}. This OH maser site was detected with a double-horned profile at 1612 MHz (first detected by \citealt{Cae1981}) and was identified as an evolved star site (\citealt{Sea1997}). Along with the double-horned profile at 1612 MHz, we also detected double-horned profiles at 1665 (first detected by \citealt{Cas1998}) and 1667 (first detected by \citealt{Dab1993}) MHz. We searched the literature and found this star is a long-period variable star, V1018 Sco (\citealt{Sae2009}). Although this star seems to be still in the AGB phase, it shows evidence for the presence of ionized gas in its circumstellar environment, as in a PN (\citealt{Coe2005}). This star was detected with non-thermal radio continuum emission (\citealt{Coe2006}), indicating that the ionization originated from shocks. Single-dish observations toward this star showed SiO  (\citealt{LN1990}; \citealt{Nye1993}) and water (\citealt{Che2017}) maser emission. Although the spatial association of these masers with V1018 Sco has not been confirmed interferometrically, their velocities are consistent with the stellar velocity determined with the OH maser spectrum. In the GLIMPSE three-color image, this star is very bright.

\noindent{\textit{G358.311$+$3.361}}. This OH maser site was detected at 1612 MHz with the Very Large Array (VLA) and was identified as an evolved star site by \citet{Zie2001}. Optical spectroscopy shows it is a post-AGB star (\citealt{Sue2006}). As well as the known 1612 MHz OH maser, we also detect two broad 1665 MHz maser spots (line widths $\sim$8\kms and $\sim$10\kms). The 1612 MHz OH maser deviates from the typical double-horned profile with five maser spots distributed over the velocity range of $-$78 to $-$45\kms. In the \textit{WISE} three-color image, this star is very bright at 12\um.

\noindent{\textit{G358.366$+$2.805}}. This OH maser site was first detected with a double-horned profile at 1612 MHz by \citet{Lie1991} and was identified as an evolved star site (\citealt{Sea1997}). Its IR variability indicates that it is in the AGB phase (\citealt{JE2006}). In the \textit{WISE} three-color image, this star is very bright at 12\um.

\noindent{\textit{G358.701$+$5.170}}. This maser site is a new detection with six 1612 MHz maser spots distributed across the velocity range of $-$190 to $-$161\kms and three 1665 MHz maser spots in the velocity range of $-$192 to $-$168\kms. Non-detections of OH emission at 1612 MHz were reported by \citet{Lie1991} and \citet{Hue1994}, with a rms of $\sim$0.09 Jy. We searched the literature and found the maser emission is associated with IRAS 17195$-$2710, which was classified as a post-AGB star based on its IR (\citealt{Hee1996}) and optical (\citealt{Sue2006}) spectra. Thus, we included this source in the evolved star category. In the \textit{WISE} three-color image, this source is associated with a bright star.

\noindent{\textit{G359.500$+$2.775}}. This maser site was detected with a single 1612 MHz spectral feature by \citet{Lie1991}, \citet{Dae1993} and \citet{Sea1997}. We find two maser spots at 1612 MHz. In the \textit{WISE} three-color image, this source is associated with a star-like object. We did not find any identification in the literature, thus this source is classified in the unknown category.

\noindent{\textit{G359.750$+$2.629}}. This OH maser site was detected with a double-horned profile at 1612 MHz in previous observations (e.g., \citealt{Bae1979}) and was identified as an evolved star site (\citealt{Sea1997}). In the \textit{WISE} three-color image, this star is very bright at 12\um. \citet{Sue2006} identified this source as a post-AGB star.

\noindent{\textit{G359.794$+$5.305}}. This maser site is a new detection with a double-horned profile at 1612 MHz and one 1667 MHz maser spot at the redshifted velocity. In the \textit{WISE} three-color image, this source is associated with a star showing bright 12\um emission and \citet{PM1988} classified this source as a possible PN. Thus, we included this source in the evolved star category. 

\noindent{\textit{G359.905$+$2.673}}. This maser site is a new detection with a double-horned profile at 1612 MHz. In the \textit{WISE} three-color image, this source is associated with a star-like object. We searched the literature and found this source is associated with a Mira-type variable star (\citealt{Mae2005}). Thus, we included this source in the evolved star category.

\noindent{\textit{G359.961$+$2.830}}. This OH maser is a new detection with only one maser spot at 1612 MHz. In the \textit{WISE} three-color image, this source is not associated with any star-like object. We did not find any identification for this source, thus included it in the unknown category.

\noindent{\textit{G359.978$+$2.563}}. This OH maser site was detected with a double-horned profile at 1612 MHz and was identified as an evolved star site (\citealt{Sea1997}). We only detected the blueshifted 1612 MHz maser component. \citet{Mae2005} and \citet{Soe2013} classified this star as a Mira-type variable star.

\noindent{\textit{G000.819$+$2.517}}. This maser site (associated with IRAS 17347$-$2653) was detected and classified as an evolved star site by \citet{Dae1993} based on the IRAS colors of the sources. We only detected the redshifted maser spot at 1612 MHz. The blueshifted maser spot at 1612 MHz in \citet{Dae1993} was $\sim$0.12 Jy, which is below our detection limit ($\sim$0.4 Jy). 

\noindent{\textit{G000.983$+$6.051}}. This maser site (associated with IRAS 17220$-$2448) was detected both by \citet{Lie1991} and \citet{Dae1993}. \citet{Dae1993} detected a double-horned profile at 1612 MHz and classified this source as an evolved star site. Like us, \citet{Lie1991} only detected the blueshifted 1612 MHz component. Thus, we included this source in the evolved star category based on the identification of \citet{Dae1993}.

\noindent{\textit{G002.081$+$3.590}}. This maser site (associated with IRAS 17337$-$2514) was detected by \citet{Lie1991} with one 1612 MHz maser spot at $+$6.1\kms. We find the 1612 MHz emission exhibits a double-horned profile, with a weaker redshifted component ($\sim$0.28 Jy) that is below the sensitivity of previous observations. In the literature, we found this source is associated with an AGB star (\citealt{Kwe1997}), thus we classified this source in the evolved star category. \citet{Che2017} reported a single-dish detection of SiO and water masers toward this source, but its association has not been confirmed interferometrically. In the \textit{WISE} three-color image, this source is associated with a bright star.

\noindent{\textit{G002.348$+$2.965}}. This OH maser site was detected with a double-horned profile at 1612 MHz and was identified as an evolved star site (\citealt{Sea1997}). In the \textit{WISE} three-color image, this star is very bright at 12\um.

\noindent{\textit{G002.388$+$2.586}}. This OH maser site (associated with IRAS 17382$-$2531) was detected both by \citet{Lie1991} and \citet{Dae1993}. \citet{Dae1993} detected the double-horned profile at 1612 MHz and classified this source as an evolved star site. \citet{Lie1991} obtained the blueshifted 1612 MHz component. We re-detected the double-horned profile at 1612 MHz. Thus, we included this source in the evolved star category based on the identification of \citet{Dae1993}. Its IR colors suggest it is a post-AGB star (\citealt{RL2009}). In the \textit{WISE} three-color image, this star is very bright at 12\um.

\noindent{\textit{G002.652$+$3.637}}. This maser site is a new detection and is associated with IRAS 17349$-$2444, which was classified as a possible PN by \citet{PM1988}. This source shows double-horned profiles at both 1612 and 1667 MHz. In the \textit{WISE} three-color image, this site is associated with a star showing bright 12\um\ emission. Thus, we identified this source as an evolved star site.

\noindent{\textit{G002.801$+$2.932 and G003.049$+$4.877}}. These two OH masers are new detections with double-horned profiles at 1612 MHz. After searching the literature, we found these two sources are associated with Mira-type variable stars (\citealt{Soe2013}). Thus, we classified them as evolved star sites.

\noindent{\textit{G003.204$+$2.642}}. This maser site (associated with IRAS 17399$-$2448) has also been detected by \citet{Lie1991} with two maser spots at velocities of $+$8.0 and $+$26.2\kms at 1612 MHz. We detected a 1612 MHz double-horned profile with two maser spots at $+$9.2 and $+$37.5\kms, which are different from the results of \citet{Lie1991}. In the \textit{WISE} three-color image, this source is associated with a bright IR star. Thus, we classified this source as an evolved star site based on the 1612 MHz spectrum and the \textit{WISE} image. Considering the mean velocities of the peaks (which should correspond to the stellar velocity) in our paper ($+$23.4\kms) and in \citeauthor{Lie1991}'s paper ($+$17.1\kms), it is possible that the emission detected by \citet{Lie1991} is from a different source.

\noindent{\textit{G003.568$+$2.652}}. This maser site is a new detection with only one maser spot at 1612 MHz. In the \textit{WISE} three-color image, this source is associated with a star-like object. After searching the literature, we found this source is associated with a Mira-type variable star (\citealt{Soe2013}). Thus, we included this source in the evolved star category.

\noindent{\textit{G003.685$+$2.160}}. This maser site (associated with IRAS 17428$-$2438) has been detected with a double-horned profile at 1612 MHz (peaked at $-$133.9 and $-$106.3\kms) and identified as an evolved star site by \citet{Lie1991}. \citet{Dae1993} also detected two 1612 MHz peaks at $-$134.90 and $-$107.58\kms. A detection at 1667 MHz was reported by \citet{Dab1993} with the velocities peaked at $-$132.78 and $-$104.69\kms. In our observations, we only detected the blueshifted 1612 MHz component. \citet{JE2006} listed this source as an AGB star based on its IRAS colors and variability.

\noindent{\textit{G003.851$+$3.534}}. This maser site (associated with IRAS 17381$-$2346) has been detected with one 1612 MHz OH maser spot at $-$1.7\kms by \citet{Lie1991}. We detected a double-horned profile at 1612 MHz, which peaks at $+$0.7 and $+$27.4\kms. In the \textit{WISE} three-color image, this source is associated with a bright IR star. Thus, we identified this source as an evolved star site based on the 1612 MHz spectrum and the \textit{WISE} image.

\noindent{\textit{G004.061$+$4.048}}. This maser site exhibits two maser spots at both 1665 and 1667 MHz and is a new detection. In the \textit{WISE} three-color image, this source is associated with a bright star-like object (IRAS 17367$-$2319), which is saturated in the center. After searching the literature, we found this source is associated with an AGB star (\citealt{Lee2001}), which is classified as a Mira-type variable star (V545 Oph) by \citet{Sae2017}. \citet{Dee2004} reported a single-dish detection of SiO masers toward this source.

\noindent{\textit{G005.885$-$0.392}}. This star formation OH maser site shows 33 maser spots at all four ground-state OH transitions and the mainline masers are a re-detection of the maser site presented in \citet{Cas1998}. This maser site was first observed at four ground-state OH transitions by \citet{Tur1969} (named as W28(A2) there) with detections at 1612, 1665 and 1667 MHz and non-detection at 1720 MHz. This star formation site is a well-known \hii region. Several papers reported water maser detections toward this site, e.g., \citet{Cae1976} and \citet{Bae1980}. A 6.7 GHz methanol maser has also been detected toward this maser site (\citealt{Cae2010}). The emission in the velocity range of $+$11 to $+$12\kms at 1720 MHz is from the nearby strong maser source G006.687$-$0.297, which is about 66 Jy at $+$11.6\kms.

\noindent{\textit{G006.095$-$0.629}}. This maser site was first detected by \citet{BK1989} and was also detected by \citet{Sea1997} with a single-peaked spectrum at 1612 MHz. We detected a double-horned profile at 1612 MHz with seven maser spots. Combining with the GLIMPSE three-color image, we classified this source as an evolved star site.

\noindent{\textit{G006.207$-$0.314}}. This maser site is a new detection with one broad maser spot at 1665 MHz (line width $\sim$12\kms) and nine maser spots at 1667 MHz. In the GLIMPSE three-color image, this source is located in the diffuse emission background. We searched the literature and found no identification for this source. Thus, we included this source in the unknown category.

\noindent{\textit{G006.584$-$0.052, G006.687$-$0.297 and G006.708$-$0.278}}. These 1720 MHz OH masers have been detected by \citet{Cle1997} as maser components three, 39 and 24 associated with the SNR W28. In the GLIMPSE three-color images, these sources are located in the diffuse emission background. The negative features in the 1720 MHz spectrum of G006.584$-$0.052 are side-lobe contamination from the nearby strong 1720 MHz maser site G006.687$-$0.297. The emission and apparent absorption features in the velocity range of $+$9 to $+$12.5\kms of G006.708$-$0.278 are also from the nearby strong source G006.687$-$0.297.

\noindent{\textit{G006.594$-$2.011}}. This OH maser site (associated with IRAS 18051$-$2415) was first detected with a 1612 MHz OH maser by \citet{Dae1993} and a 1667 MHz OH maser by \citet{Dab1993}. \citet{Sea1997} detected a double-horned profile at 1612 MHz and identified this site as an evolved star site. The IR colors of this source are consistent with a post-AGB star (\citealt{Dea2004}; \citealt{RL2009}). As well as the double-horned profile at 1612 MHz, we also obtained a double-horned spectrum at 1667 MHz. In the \textit{WISE} three-color image, this star is very bright at 12\um. 

\noindent{\textit{G006.614$-$1.902}}. This maser site (associated with MSX6C G006.6143$-$01.9021) is a new detection with a double-horned profile at 1612 MHz. Combining with the \textit{WISE} image, we classified this source as an evolved star site. \citet{Dee2005} also detected a 43 GHz SiO maser peaked at $+$31.4\kms with the Nobeyama single-dish telescope. 

\noindent{\textit{G007.121$-$0.748}}. This 1612 MHz OH maser is a new detection with a double-horned profile. Combining with the GLIMPSE image, we classified this source as an evolved star site. The spikes in the velocity range of $-$62 to $-$61\kms are RFI.

\noindent{\textit{G007.226$-$0.320}}. This 1667 MHz OH maser is a new detection with a double-horned profile. In the GLIMPSE three-color image, this source is associated with a star-like object. We found no identification in the literature, thus we included this source in the unknown category.

\noindent{\textit{G007.420$-$2.042}}. This OH maser site (associated with IRAS 18070$-$2332) was detected with a double-horned profile at 1612 MHz and was identified as an evolved star site (\citealt{Sea1997}). The IR colors of the source are consistent with a post-AGB star (\citealt{Dea2004}). Along with the double-horned profile at 1612 MHz, we also detected a double-horned profile at 1667 MHz. In the \textit{WISE} three-color image, this star is very bright at 12\um.

\noindent{\textit{G007.556$+$1.422}}. This 1612 MHz OH maser is a new detection with a double-horned profile. Combining with the \textit{WISE} image, we classified this source as an evolved star site. The spikes in the velocity range of $-$83.5 to $-$82.5\kms are RFI.

\noindent{\textit{G007.803$-$0.593}}. This 1612 MHz OH maser is a new detection with a double-horned profile. Combining with the GLIMPSE image, we classified this source as an evolved star site. The spikes in the velocity range of $-$81 to $-$80\kms are RFI.

\noindent{\textit{G007.958$-$0.393}}. This maser site has been detected by \citet{Sea1997} as a single spectral feature at 1612 MHz. We detected three maser spots at 1612 MHz. In the GLIMPSE three-color image, this source is associated with a star-like object. We did not find any identification in the literature, thus included this source in the unknown category.

\noindent{\textit{G007.961$+$1.445}}. This OH maser site was detected with a double-horned profile at 1612 MHz and was identified as an evolved star site (\citealt{Sea1997}). As well as a re-detection of the double-horned profile at 1612 MHz, we also detected a double-horned profile at 1667 MHz and one maser spot at 1665 MHz in the blueshifted velocity range. In the \textit{WISE} three-color image, this star is very bright at 12\um.

\noindent{\textit{G008.344$-$1.002}}. This maser site is associated with a well-known red supergiant -- VX Sgr (e.g., \citealt{Hue1972}). \citet{Sea1997} detected a double-horned profile at 1612 MHz toward this maser site. There are also some old detections of OH masers in this source, e.g., a 1612 MHz OH maser by \citet{CR1970} and 1612 and 1667 MHz OH masers by \citet{Pae1971}. As well as the double-horned profile at 1612 MHz, we also detected ten maser spots at 1665 MHz and ten maser spots at 1667 MHz. This maser site is the richest maser site in the evolved star category, with 36 maser spots in the 1612, 1665 and 1667 MHz transitions. In the GLIMPSE three-color image, this star is very bright and saturated in the center.

\noindent{\textit{G008.483$+$0.176}}. This OH maser site was detected with a double-horned profile at 1612 MHz and was identified as an evolved star site (\citealt{Sea1997}). \citet{Zoe1990} and \citet{Bee1994} reported a radio continuum source with flux densities of 36 mJy at 1.4 GHz and 7 mJy at 5 GHz, respectively, at a nominal position of $\simeq$4\arcsec\ away from the maser position. The spectral index of the radio continuum emission indicates a non-thermal nature. If the association between the continuum source and the maser emission is confirmed, this source could be a post-AGB star or a young PN with shock-ionized circumstellar material. We only detected the blueshifted 1612 MHz component at the peak velocity of $+$192.5\kms.

\noindent{\textit{G008.707$+$0.811}}. This OH maser site was detected with a double-horned profile at 1612 MHz and was identified as an evolved star site (\citealt{Sea1997}). We re-detected the double-horned profile at 1612 MHz, and also detected one maser spot at 1667 MHz in the redshifted velocity range. The absence of the 1667 MHz spectrum at velocities lower than $-$65\kms is due to the setup of the zoom bands described in Section \ref{observation}. In the Parkes observations, the 1667 MHz OH maser also showed a double horned profile with peaks at $\sim-$73 and $\sim-$41\kms, which is similar as the 1612 MHz OH maser. Therefore, the ATCA observations missed the blueshifted component at $\sim-$73\kms, which was probably still present. In the GLIMPSE three-color image, this star is very bright at 8\um.

\noindent{\textit{G008.733$+$0.549}}. This OH maser site was detected with a double-horned profile at 1612 MHz and was identified as an evolved star site (\citealt{Sea1997}). We re-detected the double-horned profile at 1612 MHz. In the GLIMPSE three-color image, this star is very bright at 8\um.

\noindent{\textit{G008.743$-$0.021}}. This 1667 MHz OH maser is a new detection with only one maser spot. In the GLIMPSE three-color image, it is not associated with any star-like object. We searched the literature and found no identification for this source. Thus, this site is included in the unknown category.

\noindent{\textit{G008.854$+$1.689}}. This OH maser site (associated with IRAS 17560$-$2027) was detected with a double-horned profile at 1612 MHz and was identified as an evolved star site (\citealt{Sea1997}). This source was listed as a possible PN in \citet{PM1988}. Its IR colors are consistent with a post-AGB star (\citealt{Dea2004}; \citealt{RL2009}). As well as the double-horned profile at 1612 MHz, we also detected four broad maser spots at 1665 MHz (line widths from $\sim$5 to $\sim$10\kms) and a double-horned profile at 1667 MHz. The \textit{WISE} three-color image in the bottom panel is incomplete due to the coverage of the \textit{WISE} observations.

\noindent{\textit{G008.866$-$0.249}}. This 1720 MHz OH maser is a new detection with only one maser spot. In the GLIMPSE three-color image, this source is located in the diffuse background. We searched the literature and found no identification for this source. Thus, this site is included in the unknown category.

\noindent{\textit{G008.933$-$0.015}}. This OH maser site was detected with a single-peaked spectrum at 1612 MHz (\citealt{Sea1997}). We obtained one maser spot at 1612 MHz, two maser spots at 1665 MHz and one maser spot at 1667 MHz. After searching the literature, we found this star is a red supergiant star IRC $-$ 20427 (\citealt{HB1975}) and also exhibits the 22 GHz water maser emission peaked at $+$17\kms \citep{Kle1978,Cee1988,Wae2014}. There is also a SiO detection toward this source with the single-dish telescope -- Parkes \citep{Hae1990}. In the GLIMPSE three-color image, this star is very bright and saturated in the center.

\noindent{\textit{G008.952$-$0.045}}. This OH maser site was detected with a single-peaked spectrum at 1612 MHz by \citet{Sea1997}. We obtained a double-horned profile at 1612 MHz. Combining the GLIMPSE three-color image, we classified this source as an evolved star site. 

\noindent{\textit{G009.056$-$0.156}}. This OH maser site (associated with IRAS 18033$-$2111) was first detected at 1612 MHz by \citet{Lie1991}. \citet{Sea1997} detected a double-horned profile at 1612 MHz and identified this site as an evolved star site. There is also a 1667 MHz OH maser detected by \citet{Dab1993}, which were peaked at $-$65.41 and $-$31.14\kms. The spike in the velocity range of $-$59 to $-$57\kms at 1612 MHz is from RFI.

\noindent{\textit{G009.097$-$0.392}}. This OH maser site was detected with a double-horned profile at 1612 MHz and was identified as an evolved star site by \citet{Sea1997}. This source has also been detected as OH009.1$-$0.4 (IRAS 18043$-$2116) with 1612, 1665 and 1720 MHz transitions by \citet{Seb2001}. We compared our spectra of the 1612, 1665 and 1720 MHz transitions with those of \citet{Seb2001} and found no significant variability. \citet{Seb2001} argued this star is an early post-AGB star and the 1720 MHz OH maser might trace the shock between the AGB superwind and the fast post-AGB winds. To date, the 1720 MHz OH transition has never been detected in AGB stars and is still extremely rare in other evolved objects, i.e., only one post-AGB star (this source) and three PNe (K 3$-$35, \citealt{Goe2009}; IRAS 16333$-$4807, \citealt{Qie2016a}; Vy 2$-$2, \citealt{Goe2016}). For the PNe (K 3$-$35, IRAS 16333$-$4807 and Vy 2$-$2), these 1720 MHz OH masers may trace the short-lived equatorial ejections during the formation of the PN (\citealt{Goe2009}; \citealt{Qie2016a}; \citealt{Goe2016}). Furthermore, this source is also a water fountain source with the water maser emission distributed over a velocity spread of nearly 400\kms (\citealt{Wae2009}). \citet{PS2017} also detected the radio continuum emission toward this source, which could arise from a thermal radio jet. In the GLIMIPSE three-color image, this star is very bright at 8\um. 

\begin{figure*}
\includegraphics[width=0.9\textwidth]{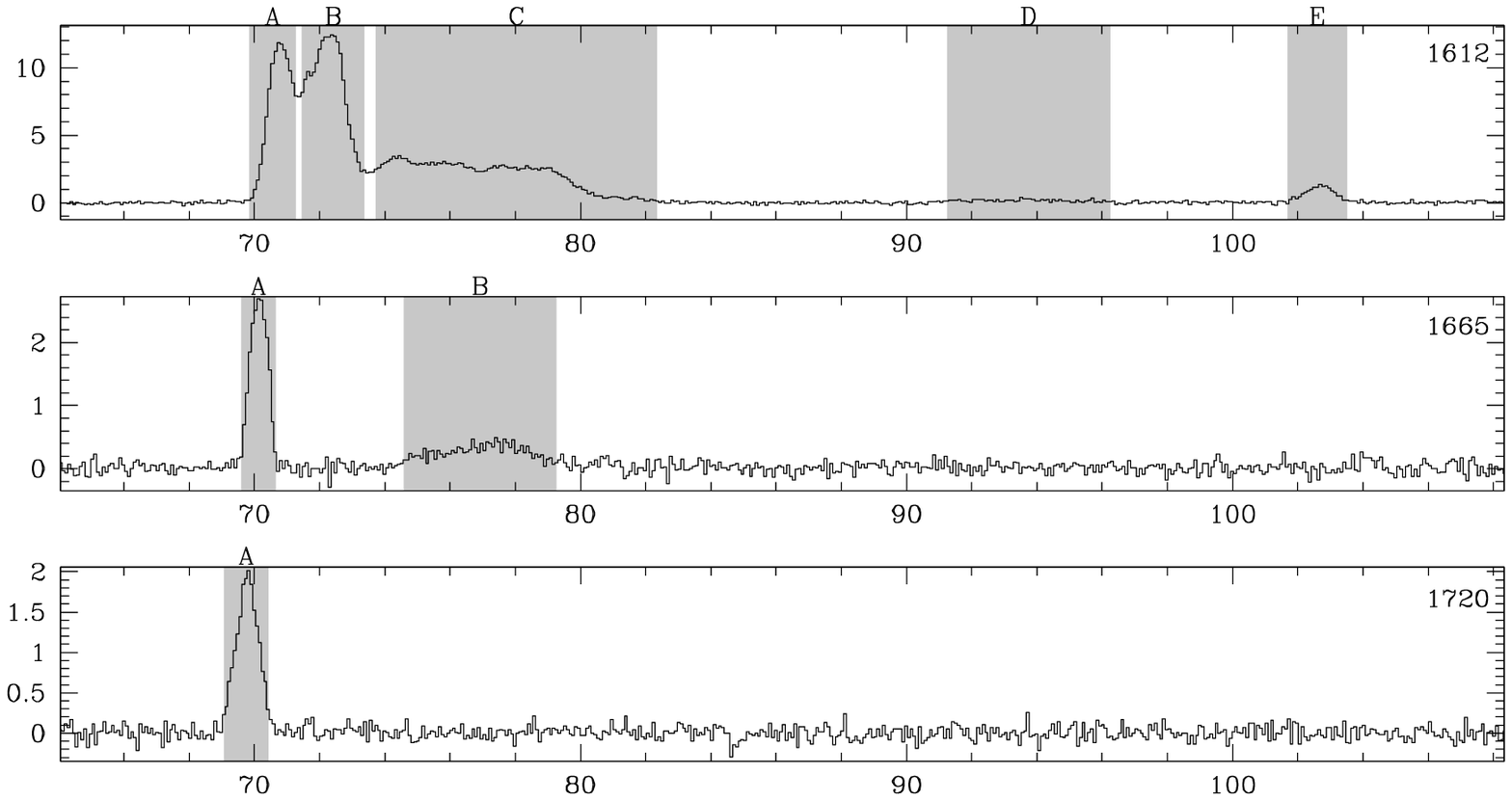}
\includegraphics[width=0.9\textwidth]{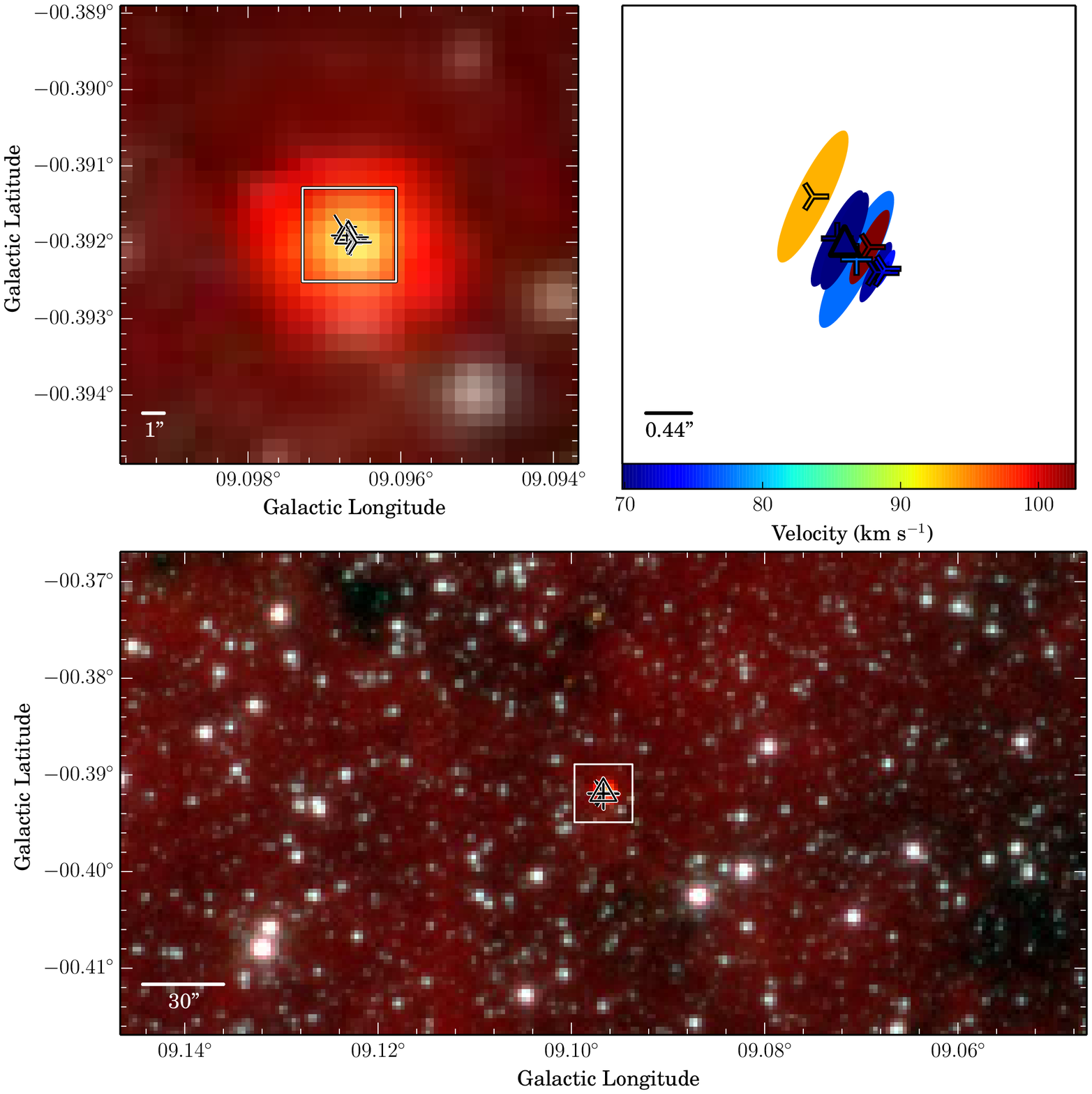}
\caption{G009.097$-$0.392 -- ES}
\label{G009.097}
\end{figure*}

\noindent{\textit{G009.539$+$0.768}}. This OH maser site (associated with IRAS 18009$-$2019) was detected with a double-horned profile at 1612 MHz and identified as an evolved star site (\citealt{Lie1989}). We obtained two maser spots at 1612 MHz, four at 1665 MHz and five at 1667 MHz. All these three transitions of OH masers were already detected by \citet{Wie1972}. In the literature, we found this star is a Mira-type variable star named V4120 Sgr (\citealt{Sae2017}). This source was first detected with 22 GHz water maser emission by \citet{Die1973} (as source IRC$-$20424). \citet{Cee1988} obtained a water maser peaked at $+$19\kms. \citet {Bae1977} detected SiO maser emission peaked at $\sim+$16\kms with Parkes. In the GLIMPSE three-color image, this star is very bright and saturated in the center.

\noindent{\textit{G009.575$-$2.032}}. This OH maser site was first detected at 1612 MHz by \citet{Lie1991}. \citet{Sea1997} detected a double-horned profile at 1612 MHz and identified it as an evolved star site. In the literature, we found this star is a Mira-type variable star (\citealt{HB2012}). A SiO maser was also detected toward this source with the single-dish telescope -- Nobeyama (\citealt{Dee2004}; \citealt{Che2017}). In the \textit{WISE} three-color image, this star is very bright and saturated in the center. The \textit{WISE} three-color image in the bottom panel is incomplete due to the coverage of the \textit{WISE} observations.

\noindent{\textit{G009.892$+$0.565}}. This 1612 MHz OH maser is a new detection with only one maser spot. In the GLIMPSE three-color image, this source is not associated with any star-like object. We searched the literature and found no identification for this source. Thus, we included this source in the evolved star category.

\noindent{\textit{G010.028$-$0.063}}. This OH maser site was first detected at 1612 MHz by \citet{Ble1994}. \citet{Sea1997} detected a double-horned profile at 1612 MHz and identified it as an evolved star site. We re-detected the double-horned profile at 1612 MHz. The GLIMPSE three-color image in the bottom panel is incomplete due to the observations of the Spitzer Space Telescope.

\section{Discussion}
\label{discussion}

\subsection{Site Categorization}
Within the Galactic regions presented in this paper, we identify 238 evolved stars with OH maser emission (66\% of all detections), 63 star formation sites (17\%), eight SNR sites and 53 unknown maser sites (15\%). Comparison with the literature (e.g., \citealt{Sea1997}; \citealt{Seb1997}; \citealt{Sea2001}; \citealt{Cas1998}; \citealt{Cas2004}) has revealed that $\sim$40\% of the evolved star sites (95/238) are new detections, $\sim$24\% of the star formation sites (15/63) are new and $\sim$94\% of the unknown maser sites (50/53) are new detections. The high number of new detections in the unknown category is consistent with previous OH maser searches being targeted towards objects with characteristics indicative of either star formation regions \citep[e.g.,][]{Ede2007} or evolved stars \citep[e.g.,][]{Lie1991}. 

The only OH emission site associated with a PN within the presented survey range is G349.509$+$1.055 (\citealt{Zie1989}), which is a re-detection of the 1612 MHz OH maser in \citet{Seb1997} but whose OH emission could be thermal (detailed in Section \ref{individual}). Some of the other evolved star (or unknown) sites may also be experiencing the PN stage, which can be further explored based on radio continuum and optical/IR spectroscopy studies. Among the eight re-detected SNR sites (\citealt{Fre1996}; \citealt{Cle1997}; \citealt{Koe1998}), five show one maser spot at 1720 MHz and three show two maser spots at 1720 MHz. These 1720 MHz OH masers trace the interaction between the SNRs and their surrounding molecular clouds (\citealt{GR1968}). A large fraction of the unknown OH masers exhibit one 1612 MHz OH maser spot and are aligned with a bright IR star-like object in the GLIMPSE or \textit{WISE} images (35/53). It is therefore likely that a large fraction of these maser sites are associated with evolved stars, but based on our identification criteria, we classified them as unknown maser sites.

In total, within the SPLASH survey (176 square degrees) covered in Paper I, Paper II and this paper, there are 933 OH maser sites, 432 (46\%) of which are new detections. Among these 933 maser sites, 629 sites are associated with evolved stars (67\%, three of which are associated with PNe), 158 sites are associated with star formation (17\%), 14 sites are associated with SNRs (2\%) and 132 sites are unknown maser sites (14\%). Most of the following Sections (\ref{size}, \ref{overlap}, \ref{evolvedstar} and \ref{starformation}) use the full 933 OH maser sites detected in the whole SPLASH survey region, thereby presenting a statistical analysis of the full sample. 

\subsection{Size of OH Maser Sites}
\label{size}

\begin{figure}
\includegraphics[width=0.4\textwidth]{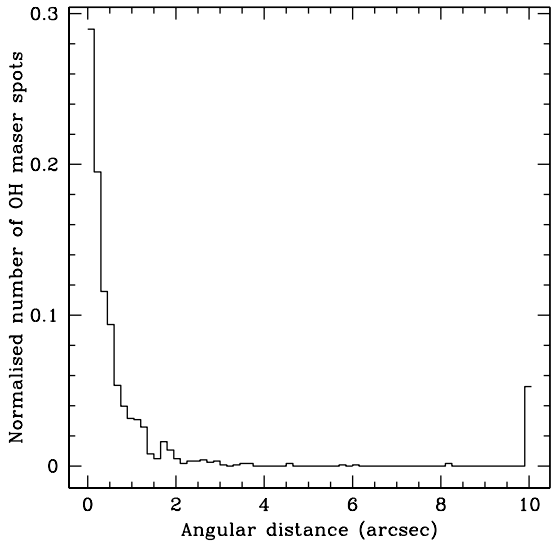}
\caption{Distribution of the angular distance to the nearest neighbour for each maser spot. This figure is cut off at 10\arcsec\ where a higher number of unrelated nearby sources begin to feature.}
\label{dis_spot}
\end{figure}

As in Paper I and Paper II, we have calculated the distribution of the separation between each maser spot and its nearest neighbour (regardless of transition; distribution is shown in Figure \ref{dis_spot}) in order to group the maser spots into sites and assess the sizes of the sites. Figure \ref{dis_spot} shows that the majority of angular separations between maser spots tend to be smaller than 2\arcsec\ and that few maser spots are separated by 2 -- 6\arcsec\ (3\%; 35/1236). We therefore adopt a source size upper limit of 4\arcsec (as in Paper I and Paper II) and consider angular separations beyond this limit on a case by case basis. In eight instances we see maser spots over more than 4\arcsec, with no clear site separation so we also consider them to be single maser sites. These include two evolved star sites (G348.668$-$0.715 and G353.751$+$0.820; both with a 4.5\arcsec\ site size), five star formation sites (G348.550$-$0.979, G348.727$-$1.037, G351.774$-$0.536, G005.885$-$0.392 and G009.621$+$0.196; site sizes of 4.6\arcsec, 5.8\arcsec, 4.4\arcsec, 4.7\arcsec\ and 10.0\arcsec, respectively) and one unknown site (G006.207$-$0.314; 4.5\arcsec). We find that there are 293 maser sites with more than one spot (227 evolved stars, 52 star formation sites, 3 SNR sites and 11 unknown sites).

\begin{figure}
\includegraphics[width=0.4\textwidth]{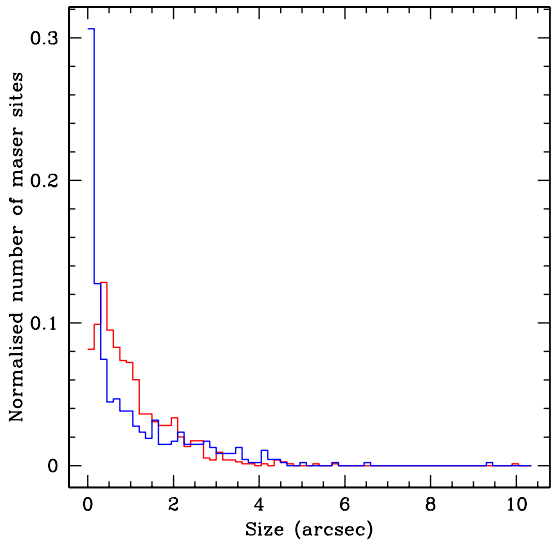}
\caption{Distribution of the sizes of 747 OH maser sites in the SPLASH survey region (red) and 470 water maser sites from HOPS (blue; \citealt{Wae2014}). All these maser sites show more than one maser spot. OH maser sites include evolved star sites, star formation sites, SNR sites and unknown sites. Water maser sites include evolved star sites, star formation sites  and unknown sites.}
\label{size_all}
\end{figure}

After adopting a maser site upper limit, we were able to calculate the sizes of maser sites using the maximum distance between OH maser spots within that site. Figure \ref{size_all} shows the site sizes of 747 OH maser sites with more than one maser spot (Paper I + Paper II + this paper; red) and the site sizes of 470 water maser sites (blue) from HOPS (which also adopted a site upper limit of 4\arcsec; \citealt{Wae2014}). The majority (97\%; 721/747) of OH maser sites in the complete SPLASH sample have site sizes smaller than 3\arcsec, which is equivalent to a linear size of about 0.12 pc at a distance of 8 kpc. Comparison between the OH and water maser site sizes in the full sample yields the same results as in Paper I: i.e., water maser sites peak at a smaller size but also include a higher number of sources with larger site sizes, and the results of a Kolmogorov-Smirnov (K-S) test shows that the OH and water maser site sizes are unlikely to be drawn from the same population (asymptotic probability $p=6\times 10^{-18}$). Paper I suggested that this is because small water maser sites are dominated by evolved stars which is associated with the circumstellar envelope closer to the star than OH masers (\citealt{RM1981}) while the larger water maser sites are predominately star formation sites (\citealt{Wae2014}) which can be associated with high-velocity outflows.

\begin{figure}
\includegraphics[width=0.4\textwidth]{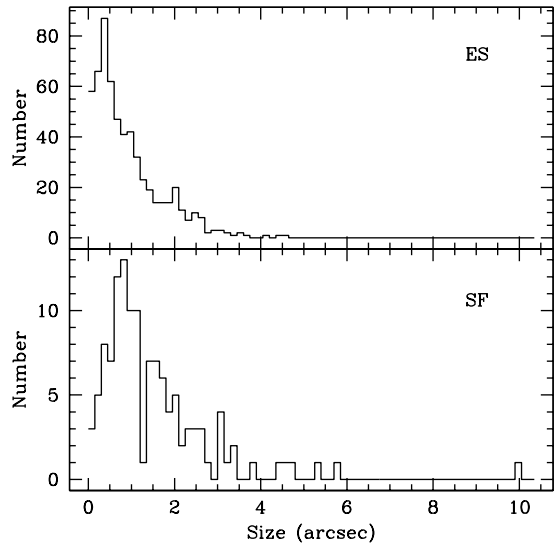}
\caption{Distribution of the sizes of 592 evolved star OH maser sites (top) and 124 star formation OH maser sites (bottom). This plot includes sources from the full SPLASH range.}
\label{size_es_sf}
\end{figure}

Figure \ref{size_es_sf} shows the distribution of OH maser site sizes from the full SPLASH sample in the categories of star formation (124 sources) and evolved stars (592 sources). We find similarly to Paper I and Paper II that the evolved stars tend to show smaller site sizes, reflecting the fact that they trace circumstellar envelopes (on scales as small as 80 au; \citealt{Rei2002}), while OH masers associated with star formation can be distributed across an associated compact \hii region (scales up to $\sim$3000 au; \citealt{FC1989}).

\subsection{Overlap between OH Transitions}
\label{overlap}
In this paper, we identified 238 evolved star maser sites. The 1612 MHz OH transition is the most prevalent in evolved star sites, with 235 of objects showing this transition, 217 of which show a typical double-horned profile. The remaining 18 sites show irregular spectra (including single spectral features), the majority of which (11) exhibit a single maser spot, with the remaining exhibiting up to ten maser spots. In comparison, there is only one evolved star site that shows a single maser spot at 1667 MHz (G344.890$+$0.120). Within the survey range of the current paper, there is only one evolved star site that shows 1720 MHz emission (G009.097$-$0.392), which is accompanied by a double-horned profile at 1612 MHz and two maser spots at 1665 MHz (discussed in Section \ref{individual}). 
Two evolved star maser sites show emission only in the main lines, with one (G352.919$+$0.063) showing a single 1667 MHz maser spot and five 1665 MHz maser spots, and the second (G004.061$+$4.048) showing two spots in each of the main lines. We note that only three evolved star sites (G344.890$+$0.120, G352.919$+$0.063 and G004.061$+$4.048) have no detected 1612 MHz maser emission.

\begin{figure}
\includegraphics[width=0.4\textwidth]{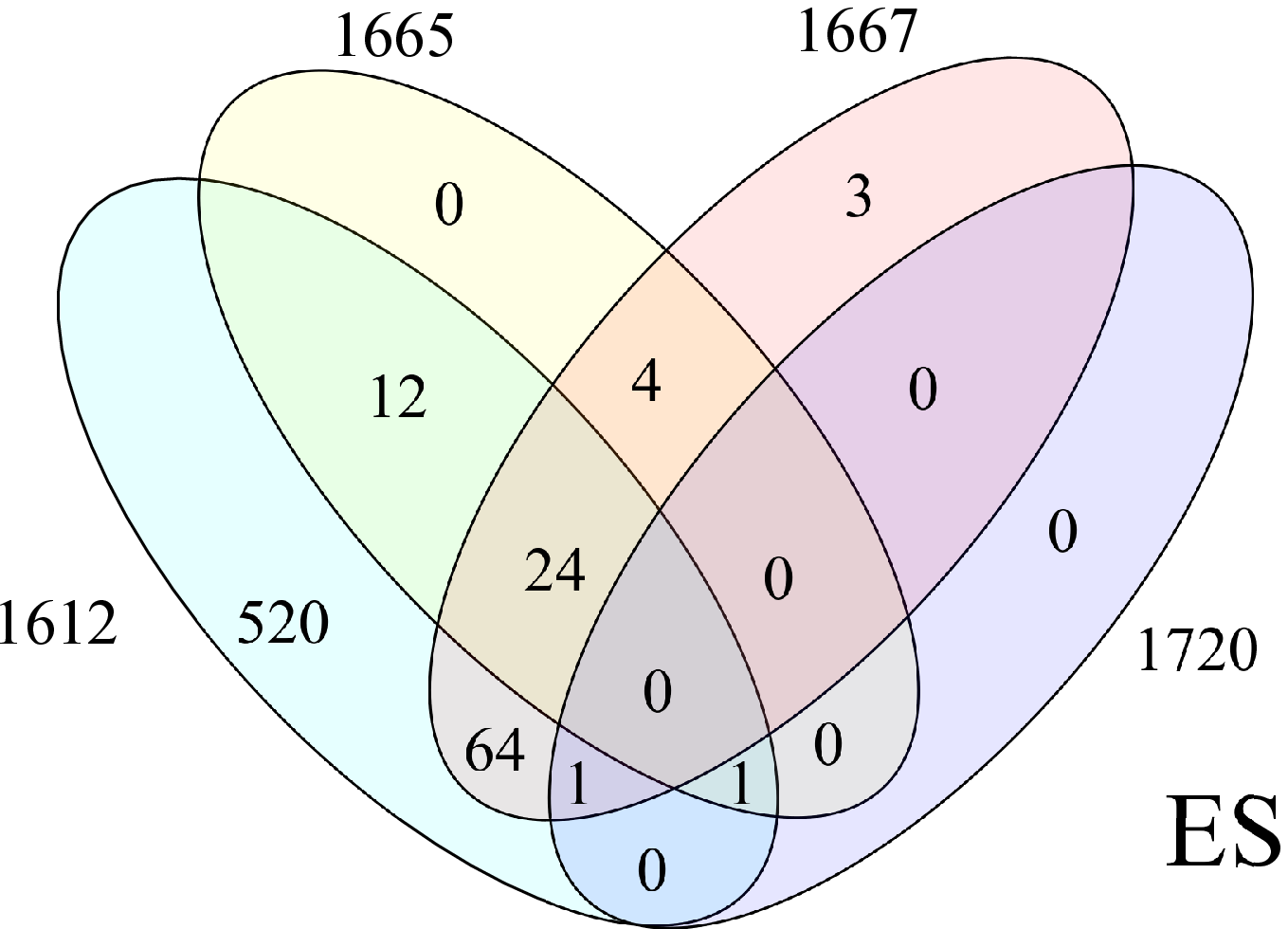}
\includegraphics[width=0.4\textwidth]{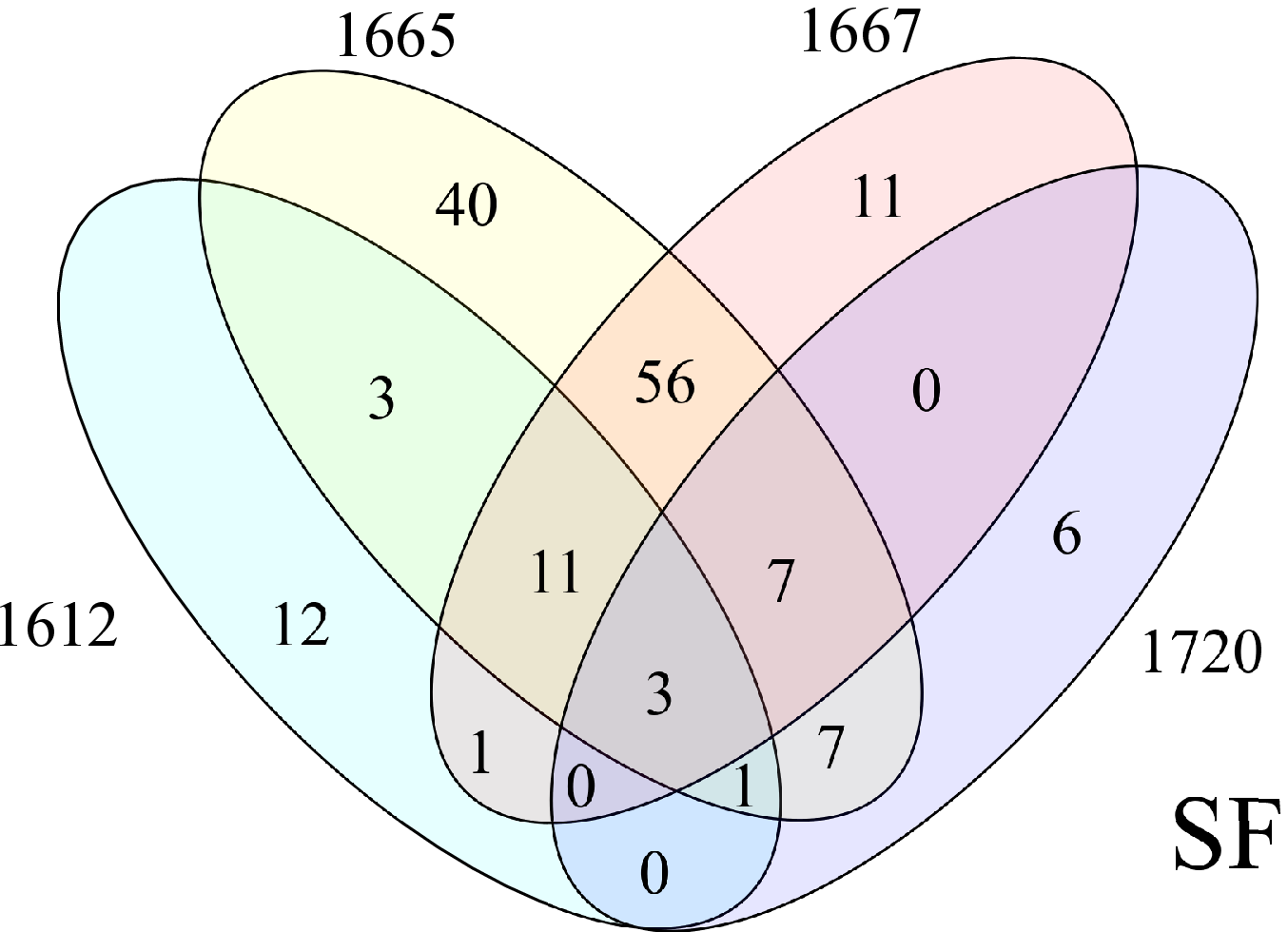}
\caption{Venn diagram of the overlap in OH transitions towards 629 evolved star (ES) OH maser sites (top panel) and 158 star formation (SF) OH maser sites (bottom panel) in the full SPLASH survey region.}
\label{vennfig}
\end{figure}

The top panel of Figure \ref{vennfig} presents a Venn diagram of the total sample of 629 evolved star sites, showing the overlap between the four ground-state OH transitions in the full SPLASH survey region. There are no solitary 1665 or 1720 MHz OH masers, but solitary 1612 MHz masers are common, and we see three instances of solitary 1667 MHz maser sources. The 1720 MHz line is only detected towards two sites: G336.644$-$0.695 (a PN; Paper I) and G009.097$-$0.392 (this paper). Overall the largest overlap is between 1612 and 1667 MHz OH masers, with 93\% (89/96) of 1667 MHz OH maser sites also exhibiting 1612 MHz detections. 1612 and 1665 MHz OH masers have the second largest overlap: 90\% (37/41) of evolved star 1665 MHz OH masers show a 1612 MHz counterpart.

The bottom panel of Figure \ref{vennfig} shows a similar Venn diagram of maser transition overlap for the full sample of 158 star formation maser sites. 1665 MHz OH masers are the most common, with the majority of star formation sites (128/158) showing 1665 MHz masers. Of these, 31\% (40/128) are solitary, i.e., do not show other ground-state OH transitions. Nevertheless, main line transitions show the largest overlap, with 87\% (77/89) of 1667 MHz masers associated with 1665 MHz OH emission, which corresponds to 60\% (77/128) of 1665 MHz masers having a 1667 MHz maser counterpart. The second-largest overlap is seen between 1665 and 1720 MHz masers, with 75\% (18/24) of 1720 MHz sources showing a 1665 MHz counterpart. The most commonly solitary transition is 1612 MHz, with 39\% (12/31) showing no association with other frequencies.  

\citet{Cas1998} detected 55 star formation OH maser sites in the region presented in the current paper, of which we re-detected 44. Among the eleven non-detections, three were reported to be weaker than 0.4 Jy at 1665 MHz, five lay in the range 0.4 -- 1.0 Jy, and the 1665 MHz flux densities of the remaining three were reported as 1.4, 1.7 and 3.0 Jy by \citet{Cas1998}. \citet{Cae2013} and \citet{Cae2014} conducted targeted observations of known OH maser sites with the Parkes telescope (observation periods: 2004 November 23 -- 27 and 2005 October 26 -- 30) to obtain full polarization spectra, including the eleven sources we failed to detect. In this later work, they re-detected ten of these sources, but did not detect the source with a peak flux density of 1.4 Jy at 1665 MHz in the earlier \citet{Cas1998} observations. Given the characteristic $\sim$0.4 Jy (5$\sigma$) detection limit of our ATCA observations, we suggest that three of the eleven non-detections are readily explained by the lower sensitivity of our observations, whereas the remaining eight are likely explained by intrinsic variability, since the time between our observations and those of \citet{Cas1998} is about 20 years. Towards the 44 OH maser sites that are common between the \citet{Cas1998} observations and our own, a similar number of sites showed both 1665 and 1667 MHz emission in the two sets of observations (29 in \citealt{Cas1998} compared with 30 in the current work). However, we detected four sources with only 1665 MHz emission in \citet{Cas1998} in both the 1665 and 1667 MHz lines, while we only detected one of the main lines towards three sites where \citet{Cas1998} found emission from both. The non-detected lines had reported peak flux densities between 0.2 and 0.7 Jy in \citet{Cas1998} and between 0.2 and 1.2 Jy in either \citet{Cae2013} or \citet{Cae2014}. Thus, these non-detections can also be accounted for by a combination of sensitivity and temporal variability.

\subsection{Evolved Star Sites}
\label{evolvedstar}

We have classified the 235 evolved star maser sites reported in this paper that have 1612 MHz OH maser emission (i.e., excluding the two evolved star sites showing only both main line transitions and the one site only with the 1667 MHz emission) according to the ratio of the integrated flux densities of their blueshifted and redshifted emission components (I$_{blue}$ and I$_{red}$). As described in Paper I and Paper II, we may use the integrated flux densities to estimate the number of photons originating from each side of the circumstellar envelope, and adopt the following criterion for classification: if the ratio of I$_{blue}$ and I$_{red}$ is between 0.5 and 2, we categorize the source as symmetric, otherwise, we categorize the source as asymmetric. This latter category includes the 11 sources which only exhibit one 1612 MHz maser spot, for which the ratio is satisfied by the integrated flux density of the detected feature and the 5$\sigma$ detection limit. Within the Galactic region newly presented in this work, 170/235 (72\%) sources are symmetric and 65/235 (28\%) are classified as asymmetric. As described in Paper I and Paper II, we have obtained the mid-IR (\textit{WISE}) counterparts for sources within these two categories, since emission from circumstellar dust can dominate compared to photospheric radiation for wavelengths above 5\um  (\citealt{Ble2006}). \textit{WISE} [12]-[22] (12\um and 22\um)  colors were obtained for the mid-IR counterparts of 155 (91\%) symmetric sources and 55 (85\%) asymmetric sources.

\begin{figure}
\includegraphics[width=0.4\textwidth]{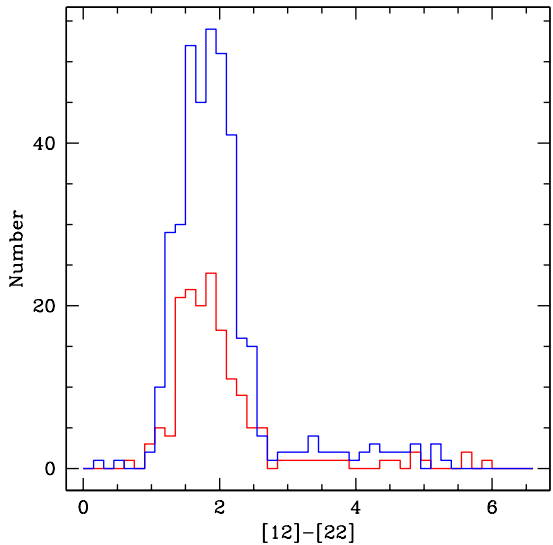}
\caption{\textit{WISE} [12]-[22] color distributions of 387 symmetric (blue) and 162 asymmetric (red) evolved star sites showing 1612 MHz transition in the full SPLASH survey region.}
\label{hist_1222}
\end{figure}

Combined with the same information from the Galactic regions given in Paper I and Paper II, there are a total of 387 symmetric sources and 162 asymmetric sources with \textit{WISE} [12]-[22] colors. When considering the full sample, we find no obvious difference in the \textit{WISE} [12]-[22] color of the two categories, in contrast to the results of Paper I. Histograms of the two distributions are shown in Figure \ref{hist_1222}. A K-S test yields $p=0.68$ for the full sample (compared to $p=0.1$ for the subsample in Paper I), consistent with both distributions being drawn from the same underlying population. This is consistent with the findings of Paper II.

\begin{figure}
\includegraphics[width=0.4\textwidth]{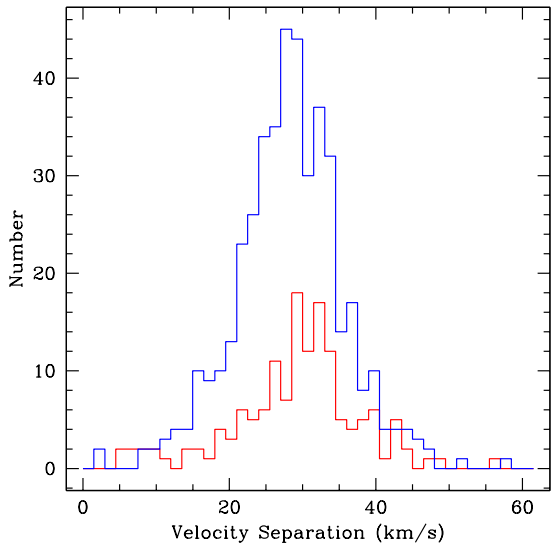}
\caption{The velocity separation $\bigtriangleup V$ of 433 symmetric (blue) and 146 asymmetric (red) evolved star sites showing 1612 MHz transition in the SPLASH survey region.}
\label{Vel_Sep}
\end{figure}


Figure \ref{Vel_Sep} illustrates the velocity separation of the
most extreme spectral features associated with symmetric (433) and
asymmetric (146) sources in the full
SPLASH survey region. Although the distribution largely overlap, they
seem qualitatively different, with a trend of asymmetric sources
appearing to have larger velocity separations than symmetric sources.
A K-S test shows $p=0.001$, supporting the idea that the populations
from which the two distributions are drawn are genuinely different. This
suggests that symmetric and asymmetric
evolved star OH maser sites have similar mid-IR color
properties, but different velocity separation properties.
We also tested the median values of the velocity separations (28.30 and
29.87 \kms for symmetric and asymmetric sources, respectively)
with a Mood's median test. It suggests that the medians of the two
distributions do differ ($p=0.046$).
According to \citet{BH1983},
the expansion velocities of OH/IR stars are faster for
higher main-sequence masses. Since the velocity separation of masers is
a direct indication of the  expansion velocity, our results would
suggest that the asymmetric sources may have higher
main-sequence masses than the symmetric sources. While this qualitative
trend may by present, a difference in initial mass is, however,
difficult to quantify, since expansion velocities also depend on other
parameters, such as evolutionary stage along the AGB \citep{Vassiliadis93} or metallicity \citep{Goldman17}.
Moreover, we may need to further consider potential biases
such as the value of the $I_{blue}$ vs. $I_{red}$ ratio that we chose
to separate the two categories.

\subsection{Star Formation Sites}
\label{starformation}
\begin{figure}
\includegraphics[width=0.4\textwidth]{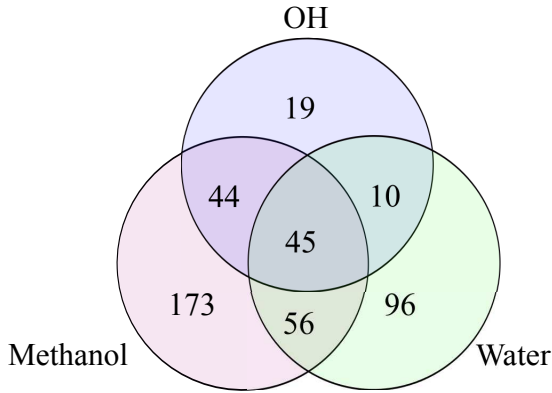}
\caption{Venn diagram showing the overlap of OH, 6.7 GHz methanol and 22 GHz water masers from star formation regions, between Galactic longitudes of $332^{\circ}$ and $10^{\circ}$ and Galactic latitudes of $\pm0.5^{\circ}$.}
\label{vennsf}
\end{figure}

We have compared star formation OH maser sites in the SPLASH survey region to 6.7 GHz methanol maser sites from the MMB survey (\citealt{Cae2010}; \citealt{Gre2010}; \citealt{Cae2011}) and 22 GHz water maser sites from HOPS (\citealt{Wae2011}; \citealt{Wae2014}). The three surveys overlap between of $332^{\circ} < l < 10^{\circ}$ and $b \pm0.5^{\circ}$; this region contains 118 star formation OH maser sites (1$\sigma$ sensitivity of $\sim$0.07 Jy; Paper I, Paper II and this paper), 318 methanol maser sites (1$\sigma$ sensitivity of $\sim$0.17 Jy; \citealt{Gre2009}), and 207 star formation water maser sites (1$\sigma$ sensitivity of $\sim$1 $-$ 2 Jy; \citealt{Wae2011}). The association between these three maser species is shown in Figure \ref{vennsf}. As in Paper I and Paper II, we used an association threshold of 5.1\arcsec\ (0.001\degree in both longitude and latitude) to determine associations between all species. For OH maser sites, the largest overlap is with methanol maser sites, with 89/118 (75\%) OH maser sites having associated methanol emission. This is similar to the values found in Paper I (73\%) and Paper II (74\%). 55 OH maser sites (47\%) also show 22 GHz water masers -- also a comparable fraction to that found in Paper I (the 44\% association rate) and Paper II (37\% association rate). Water masers also have the largest overlap with methanol masers, with 101/207 (49\%) water masers associated with methanol maser sites. This is slightly lower than the value found in Paper I (57\%) and is slightly higher than the ratio (40\%) found in \citet{Bre2018}. The region of \citet{Bre2018} is larger than ours (and they use a slightly different association thresholds), which is between Galactic longitudes of $290^{\circ}$ and $30^{\circ}$ and Galactic latitudes of $\pm0.5^{\circ}$. 173/318 (54\%) methanol maser sites are not associated with either OH or water masers. This is higher than the fractions of solitary OH (16\%) and solitary water maser sites (46\%). However, as discussed in Papers I and II, the association fractions between OH/methanol and water masers are affected by the low sensitivity of HOPS, which has a typical rms noise of about 1 $-$ 2 Jy \citep{Wae2011}. 
Therefore, including \citet{Wae2014}, we also compared our OH masers with water masers from the sensitive surveys of \citet{Bre2010}, \citet{Tie2014} and \citet{Tie2016}. These studies were targeted towards OH and methanol masers, and had typical rms noise values of 0.1 Jy. We found that 80/118 (68\%) OH masers are associated with water masers, which is significantly lower than the value found in \citet{Bre2010} (79\%). As described above, \citet{Bre2010} is targeted towards a sample of OH and methanol masers that were detected in both targeted and full survey observations themselves. Thus, it is possible that their association rate is biased. Furthermore, sensitive water maser observations are yet to be conducted towards a number of the OH masers within the SPLASH region, including the 41 (out of 118; 35\%) star formation OH masers that are new detections. Therefore, the association rate (68\%) between the SPLASH star formation OH masers and water masers \citep{Bre2010,Tie2014,Tie2016,Wae2014} should be considered to be a lower limit. Given that both OH masers and water masers are variable (e.g., \citealt{Bre2010} and \citealt{Cae2014}), the lower association rate might be also partly due to the variability of these two species of masers. 

As in Paper I and Paper II, we also utilized the 1720 MHz zoom band to investigate the 1.7 GHz radio continuum properties of the 63 star formation OH maser sites newly repored in this work (These observations had a  continuum bandwidth of 2 MHz during the 2015 OCT Semester and 2.5 MHz during the 2014 OCT Semester). We found that 16/63 (25\%) star formation OH maser sites are associated with radio continuum sources at 1.7 GHz (with a continuum rms noise typically $\sim$10 mJy), which is higher than the detection rates in Paper I (9\%) and Paper II (19\%). In the full SPLASH survey region, the total detection rate of radio continuum sources at 1.7 GHz is 18\% (28/158), which is lower than the 38\% radio continuum source association rate (for continuum observations at 8.2 and 9.2 GHz) reported by \citet{FC2000} towards a sample of star formation OH masers (the typical continuum rms noise was about 0.15 mJy). As discussed in Paper I and Paper II, the low continuum association rate in the SPLASH survey region may be due to a combination of the low frequency (1.7 GHz) and limited sensitivity of our zoom band continuum observations.

\subsection{OH maser candidates not detected in the ATCA observations}
\label{nodetections}

There were 17 OH maser candidates identified in the Parkes observations towards which we did not detect any OH maser emission in the present work. These are listed in Table \ref{nondetection}. Of these, 7/17 (marked with a ``Z'' in the table) have velocities that fell outside the frequency range of our ATCA zoom bands. 2/17 candidates (G345.15$-$1.00 and G354.30$-$0.15) were not observed with the ATCA; G345.15$-$1.00 was incorrectly categorized as thermal emission from the Parkes observations, and G354.30$-$0.15 was accidentally omitted from our target list. We checked a nearby target field (G354.650$+$0.00; $\sim$23\arcmin\ away), but did not find the weak OH maser emission detected in the Parkes observations. This is consitent with the fact that the peak flux densities of these maser candidates were 0.44 Jy at 1665 MHz and 0.39 Jy at 1667 MHz during the Parkes observations, and the detection limit at the offset position in the ATCA observations was $\sim$0.8 Jy. One site (1/17; G349.50$+$1.05) is coincident with a PN (\citealt{Zie1989}), but its 1612 MHz emission could be of thermal nature, as detailed in Section \ref{individual}.

7/17 of the remaining undetected maser sites (marked with a ``W'' in Table \ref{nondetection}) were very weak in the Parkes observations (six were less than 0.22 Jy and one was 0.42 Jy). Of these, three (3/7) had double-horned profiles at 1612 MHz (typical of evolved stars), three (3/7) exhibited a single 1612 MHz spectral feature and one (1/7) had a single spectral feature in the 1667 MHz transition. It is likely that these seven sources were undetected simply due to inadequate sensitivity in the followup observations presented here, which typically had a 5$\sigma$ detection limit of around 0.4 Jy. This would be particularly true if they have experienced any reduction flux density in the 3 years between the Parkes and ATCA observations. Such temporal variability is fairly common in ground-state OH masers. Indeed, \citet{Cae2014} reported that fewer than 10\% of a sample of 187 ground-state OH masers remained stable over a period of decades.

\tabletypesize{\small}
\begin{table}
\begin{center}
\caption{\textnormal{List of positions which exhibit maser emission in the Parkes observations, but were not detected in our ATCA observations.}}
\label{nondetection}
\begin{tabular}{ll}
\hline
\hline
G332.275$-$0.05(A,Z)&G345.15$-$1.00(A,W,N)\\
G348.45$+$1.65(A,W)&G349.50$+$1.05(T)\\
G350.10$+$0.75(D,Z)&G350.35$-$1.50(D,Z)\\
G351.05$+$2.05(D,Z)&G352.35$-$0.55(A,W)\\
G352.55$-$2.10(A,W)&G353.65$+$0.80(D,Z)\\
G353.75$-$1.55(D,Z)&G353.95$-$1.00(D,Z)\\
G354.30$-$0.15(D,N)&G354.40$+$1.80(A,W)\\
G358.75$+$2.85(D,W)&G007.30$+$1.20(D,W)\\
G007.70$+$0.70(D,W)\\

\hline
\end{tabular}
\end{center}

\textbf{Notes.} D -- a double-horned spectrum; A -- a single-peaked spectrum; W -- weak ($<$0.42 Jy) in Parkes observations; Z -- missed due to the coverage of zoom bands; N -- not observed; T -- thermal emission.

\end{table}

\section{Conclusions}

In this paper, we present accurate positions for OH masers at all four ground-state transitions, within the remainder of the SPLASH survey region, which covers Galactic longitudes of 332\degree and 334\degree and Galactic latitudes of $-$2\degree and $+$2\degree, between Galactic longitudes of 344\degree and 355\degree and Galactic latitudes of $-$2\degree and $+$2\degree, between Galactic longitudes of 358\degree and 4\degree and Galactic latitudes of $+$2\degree and $+$6\degree, and between Galactic longitudes of 5\degree and 10\degree and Galactic latitudes of $-$2\degree and $+$2\degree (see Figure \ref{glb}). We detect a total of 362 maser sites, which show emission in one, two, three or four of the ground-state OH transitions. About half of the reported maser sites (160/362) are new detections, seen for the first time in the Parkes SPLASH observations, and confirmed in the present work with the ATCA. The improved resolution of the interferometric data presented here allows us to identify astrophysical objects associated with these maser sites, and we have categorised them as either evolved stars, star formation regions, SNRs or unknown.

238 OH maser sites are associated with evolved stars. These usually exhibit characteristic double-horned profiles at 1612 MHz, and are occasionally accompanied by 1665 and/or 1667 MHz OH counterparts. One rare evolved star site (G009.097$-$0.392) not only shows the 1612 and 1665 MHz transitions, but also has a 1720 MHz OH maser, which might trace the shock between the AGB superwind and the fast post-AGB winds. 63 maser sites are associated with star formation. These commonly show multiple strong maser spots in the main line transitions and occasionally also contain 1612 or 1720 MHz OH masers. Of the remaining maser sites, eight are categorized as associated with SNRs. The remaining 53 maser sites are classified as ``unknown'', since we are unable make a positive association from the literature and IR images. However, we suggest that a large fraction of these are likely to be associated with evolved stars. 

Statistical analysis is conducted on the OH maser sites in the full SPLASH survey region (176 square degrees), which combines the data from Paper I, Paper II and the current paper. Based on the accurate positions obtained in this work, the angular sizes of most OH maser sites (97\%) are smaller than 3$\arcsec$. Evolved star OH maser sites in the survey region show smaller sizes than star formation OH maser sites (as was found in Papers I and II).

We classify evolved star sites in the full SPLASH survey region according to the integrated 1612 MHz flux densities of their blueshifted and redshifted components. In Paper I, a K-S test suggested that the distributions of \textit{WISE} colors of symmetric and asymmetric sources might be different ($p=0.1$), but this is not confirmed in the full SPLASH sample ($p=0.68$). With the much larger sample presented here (the full SPLASH survey), we now find that the velocity separations of asymmetric sources seem to be larger than those of symmetric sources, which was not apparent in the samples presented in either Paper I or II alone. From this, we suggest that asymmetric sources may have higher main-sequence masses than symmetric sources.


We find that 28 of the 158 star formation OH maser sites detected in the full SPLASH survey region are associated with 1.7 GHz continuum sources, which is lower than the ratio of OH maser sites associated with 8.2 and 9.2 GHz continuum sources (\citealt{FC2000}). The difference in detection rates is likely explained by our lower observing frequency combined with the relatively low sensitivity of our continuum observations, but may also be contributed to by the fact that the observations of \cite{FC2000} were targeted towards 26 fields containing known OH and water masers \citep[a sample that itself originated from a partially-targeted OH maser sample,][]{CH1983b}.

17 of the target fields newly presented in this paper do not contain any detected maser emission. Seven of them are missed due to the velocity coverage of the zoom bands. From the Parkes spectra, we find that further seven sources have weak spectral profiles, and their non-detections can easily be explained by inadequate sensitivity, possibly combined with temporal variability.

\acknowledgments The Australia Telescope Compact Array is part of the Australia Telescope which is funded by the Commonwealth of Australia for operation as a National Facility managed by CSIRO. This research has made use of: NASA's Astrophysics Data System Abstract Service; and the SIMBAD data base, operated at CDS, Strasbourg, France. This work is based in part on observations made with the Spitzer Space Telescope, which is operated by the Jet Propulsion Laboratory, California Institute of Technology under a contract with National Aeronautics and Space Administration (NASA). This publication also makes use of data products from the Wide-field Infrared Survey Explorer, which is a joint project of the University of California Institute of Technology, funded by NASA. H.-H.Q. is partially supported by the Special Funding for Advanced Users, budgeted and administrated by Center for Astronomical Mega-Science, Chinese Academy of Sciences (CAMS-CAS), CAS ``Light of West China'' Program and the National Natural Science Foundation of China (Grant No. 11903038). JFG is partially supported by MINECO (Spain) grant AYA2017-84390-C2-1-R (co-funded by FEDER) and by the State Agency for Research of the Spanish MCIU through the ``Center of Excellence Severo Ochoa'' award for the Instituto de Astrof\'{\i}sica de Andaluc\'{\i}a (SEV-2017-0709). This work was supported in part by the Major Program of the National Natural Science Foundation of China (Grant Nos. 11590780, 11590784) and the CAS Pioneer Hundred Talents Program(Technological excellence, Y650YC1201). HI was supported by the JSPS Bilateral Collaboration Program and KAKENHI programs 25610043 and 16H02167.

\clearpage


\label{lastpage}


\begin{thebibliography}{99}

\bibitem[Argon et al.(2000)]
{Are2000} Argon, A.~L., Reid, M.~J., \& Menten, K.~M.\ 2000, \apjs, 129, 159 

\bibitem[Baan et al.(1982)]
{Bae1982} Baan, W.~A., Wood, P.~A.~D., \& Haschick, A.~D.\ 1982, \apjl, 260, L49 

\bibitem[Balister et al.(1977)]
{Bae1977} Balister, M., Batchelor, R.~A., Haynes, R.~F., et al.\ 1977, \mnras, 180, 415

\bibitem[Batchelor et al.(1980)]
{Bae1980} Batchelor, R.~A., Caswell, J.~L., Goss, W.~M., et al.\ 1980, Australian Journal of Physics, 33, 139

\bibitem[Baud et al.(1979)]
{Bae1979} Baud, B., Habing, H.~J., Matthews, H.~E., et al.\ 1979, \aaps, 35, 179

\bibitem[Baud, \& Habing(1983)]
{BH1983} Baud, B., \& Habing, H.~J.\ 1983, \aap, 127, 73

\bibitem[Becker et al.(1994)]
{Bee1994} Becker, R.~H., White, R.~L., Helfand, D.~J., et al.\ 1994, \apjs, 91, 347

\bibitem[Blommaert et al.(1994)]
{Ble1994} Blommaert, J.~A.~D.~L., van Langevelde, H.~J., \& Michiels, W.~F.~P.\ 1994, \aap, 287, 479

\bibitem[Blum et al.(2006)]
{Ble2006} Blum, R.~D., Mould, J.~R., Olsen, K.~A., et al.\ 2006, \aj, 132, 2034 

\bibitem[Bowers \& Knapp(1989)]
{BK1989} Bowers, P.~F., \& Knapp, G.~R.\ 1989, \apj, 347, 325 

\bibitem[Breen et al.(2010)]
{Bre2010} Breen, S.~L., Caswell, J.~L., Ellingsen, S.~P., \& Phillips, C.~J.\ 2010, \mnras, 406, 1487 

\bibitem[Breen et al.(2013)]
{Bre2013} Breen, S.~L., Ellingsen, S.~P., Contreras, Y., et al.\ 2013, \mnras, 435, 524 

\bibitem[Breen et al.(2018)]
{Bre2018} Breen, S.~L., Contreras, Y., Ellingsen, S.~P., et al.\ 2018, \mnras, 474, 3898 

\bibitem[Caswell, \& Robinson(1970)]
{CR1970} Caswell, J.~L., \& Robinson, B.~J.\ 1970, \aplett, 7, 75

\bibitem[Caswell et al.(1976)]
{Cae1976} Caswell, J.~L., Batchelor, R.~A., Goss, W.~M., et al.\ 1976, Proceedings of the Astronomical Society of Australia, 3, 61

\bibitem[Caswell et al.(1980)]
{Cae1980} Caswell, J.~L., Haynes, R.~F., \& Goss, W.~M.\ 1980, Australian Journal of Physics, 33, 639 

\bibitem[Caswell et al.(1981)]
{Cae1981} Caswell, J.~L., Haynes, R.~F., Goss, W.~M., et al.\ 1981, Australian Journal of Physics, 34, 333

\bibitem[Caswell \& Haynes(1983a)]
{CH1983a} Caswell, J.~L., \& Haynes, R.~F.\ 1983a, Australian Journal of Physics, 36, 361 

\bibitem[Caswell \& Haynes(1983b)]
{CH1983b} Caswell, J.~L., \& Haynes, R.~F.\ 1983b, Australian Journal of Physics, 36, 417 

\bibitem[Caswell \& Haynes(1987)]
{CH1987} Caswell, J.~L., \& Haynes, R.~F.\ 1987, Australian Journal of Physics, 40, 215 

\bibitem[Caswell(1998)]
{Cas1998} Caswell, J.~L.\ 1998, \mnras, 297, 215

\bibitem[Caswell(1999)]
{Cas1999} Caswell, J.~L.\ 1999, \mnras, 308, 683 

\bibitem[Caswell(2004)]
{Cas2004} Caswell, J.~L.\ 2004, \mnras, 349, 99  

\bibitem[Caswell et al.(2010)]
{Cae2010} Caswell, J.~L., Fuller, G.~A., Green, J.~A., et al.\ 2010, \mnras, 404, 1029 

\bibitem[Caswell et al.(2011)]
{Cae2011} Caswell, J.~L., Fuller, G.~A., Green, J.~A., et al.\ 2011, \mnras, 417, 1964 

\bibitem[Caswell et al.(2013)]
{Cae2013} Caswell, J.~L., Green, J.~A., \& Phillips, C.~J.\ 2013, \mnras, 431, 1180 

\bibitem[Caswell et al.(2014)]
{Cae2014} Caswell, J.~L., Green, J.~A., \& Phillips, C.~J.\ 2014, \mnras, 439, 1680 

\bibitem[Cesaroni et al.(1988)]
{Cee1988} Cesaroni, R., Palagi, F., Felli, M., et al.\ 1988, \aaps, 76, 445 

\bibitem[Cho et al.(2017)]
{Che2017} Cho, C.-Y., Cho, S.-H., Kim, S., et al.\ 2017, \apjs, 232, 13

\bibitem[Claussen et al.(1997)]
{Cle1997} Claussen, M.~J., Frail, D.~A., Goss, W.~M., \& Gaume, R.~A.\ 1997, \apj, 489, 143 

\bibitem[Cohen et al.(1995)]
{Coe1995} Cohen, R.~J., Masheder, M.~R.~W., \& Caswell, J.~L.\ 1995, \mnras, 274, 808 

\bibitem[Cohen et al.(2005)]
{Coe2005} Cohen, M., Parker, Q.~A., \& Chapman, J.\ 2005, \mnras, 357, 1189

\bibitem[Cohen et al.(2006)]
{Coe2006} Cohen, M., Chapman, J.~M., Deacon, R.~M., et al.\ 2006, \mnras, 369, 189

\bibitem[Cyganowski et al.(2008)]
{Cye2008} Cyganowski, C.~J., Whitney, B.~A., Holden, E., et al.\ 2008, \aj, 136, 2391-2412

\bibitem[David et al.(1993a)]
{Dae1993} David, P., Le Squeren, A.~M., \& Sivagnanam, P.\ 1993a, \aap, 277, 453

\bibitem[David et al.(1993b)]
{Dab1993} David, P., Le Squeren, A.~M., Sivagnanam, P., et al.\ 1993b, \aaps, 98, 245 

\bibitem[Dawson et al.(2014)]
{Dae2014} Dawson, J.~R., Walsh, A.~J., Jones, P.~A., et al.\ 2014, \mnras, 439, 1596 

\bibitem[Deacon et al.(2004)]
{Dea2004} Deacon, R.~M., Chapman, J.~M., \& Green, A.~J.\ 2004, \apjs, 155, 595

\bibitem[Deacon et al.(2007)]
{Dee2007} Deacon, R.~M., Chapman, J.~M., Green, A.~J., \& Sevenster, M.~N.\ 2007, \apj, 658, 1096 

\bibitem[Deguchi et al.(2000)]
{Dee2000} Deguchi, S., Fujii, T., Izumiura, H., et al.\ 2000, \apjs, 130, 351 

\bibitem[Deguchi et al.(2004)]
{Dee2004} Deguchi, S., Fujii, T., Glass, I.~S., et al.\ 2004, \pasj, 56, 765 

\bibitem[Deguchi et al.(2005)]
{Dee2005} Deguchi, S., Nakashima, J.-I., Miyata, T., \& Ita, Y.\ 2005, \pasj, 57, 933 

\bibitem[Derue et al.(2002)]
{Dee2002} Derue, F., Marquette, J.-B., Lupone, S., et al.\ 2002, \aap, 389, 149

\bibitem[Dickinson et al.(1973)]
{Die1973} Dickinson, D.~F., Bechis, K.~P., \& Barrett, A.~H.\ 1973, \apj, 180, 831 

\bibitem[Dodson \& Ellingsen(2002)]
{DE2002} Dodson, R.~G., \& Ellingsen, S.~P.\ 2002, \mnras, 333, 307 

\bibitem[Edris et al.(2007)]
{Ede2007} Edris, K.~A., Fuller, G.~A., \& Cohen, R.~J.\ 2007, \aap, 465, 865 

\bibitem[Forster \& Caswell(1989)]
{FC1989} Forster, J.~R., \& Caswell, J.~L.\ 1989, \aap, 213, 339 

\bibitem[Forster \& Caswell(2000)]
{FC2000} Forster, J.~R., \& Caswell, J.~L.\ 2000, \apj, 530, 371 

\bibitem[Frail et al.(1996)]
{Fre1996} Frail, D.~A., Goss, W.~M., Reynoso, E.~M., et al.\ 1996, \aj, 111, 1651 

\bibitem[Garc{\'\i}a-Hern{\'a}ndez et al.(2007)]
{GHe2007} Garc{\'\i}a-Hern{\'a}ndez, D.~A., Perea-Calder{\'o}n, J.~V., Bobrowsky, M., et al.\ 2007, \apjl, 666, L33

\bibitem[G{\'e}rard et al.(1998)]
{Gee1998} G{\'e}rard, E., Crovisier, J., Colom, P., et al.\ 1998, \planss, 46, 569 

\bibitem[Goedhart et al.(2000)]
{Goe2000} Goedhart, S., van der Walt, D.~J., \& Schutte, A.~J.\ 2000, \mnras, 315, 316 

\bibitem[Goldman et al.(2017)]
{Goldman17} Goldman, S. R., et al.\ 2017, \mnras, 465, 403 

\bibitem[G{\'o}mez et al.(2009)]
{Goe2009} G{\'o}mez, Y., Tafoya, D., Anglada, G., et al.\ 2009, \apj, 695, 930 

\bibitem[G{\'o}mez et al.(2016)]
{Goe2016} G{\'o}mez, J.~F., Uscanga, L., Green, J.~A., et al.\ 2016, \mnras, 461, 3259 

\bibitem[Goss \& Robinson(1968)]
{GR1968} Goss, W.~M., \& Robinson, B.~J.\ 1968, \aplett, 2, 81 

\bibitem[Green et al.(2009)]
{Gre2009} Green, J.~A., Caswell, J.~L., Fuller, G.~A., et al.\ 2009, \mnras, 392, 783 

\bibitem[Green et al.(2010)]
{Gre2010} Green, J.~A., Caswell, J.~L., Fuller, G.~A., et al.\ 2010, \mnras, 409, 913 

\bibitem[Gundermann(1965)]
{Gun1965} Gundermann, E.~J.\ 1965, Ph.D.~Thesis,  

\bibitem[Hall et al.(1990)]
{Hae1990} Hall, P.~J., Wright, A.~E., Troup, E.~R., et al.\ 1990, \mnras, 247, 549

\bibitem[Hansen \& Blanco(1975)]
{HB1975} Hansen, O.~L., \& Blanco, V.~M.\ 1975, \aj, 80, 1011 

\bibitem[Hashimoto(1994)]
{Has1994} Hashimoto, O.\ 1994, \aaps, 107, 445

\bibitem[Henning et al.(1996)]
{Hee1996} Henning, T., Chan, S.~J., \& Assendorp, R.\ 1996, \aap, 312, 511

\bibitem[Hu et al.(1994)]
{Hue1994} Hu, J.~Y., te Lintel Hekkert, P., Slijkhuis, F., et al.\ 1994, \aaps, 103, 301

\bibitem[Huemmerich \& Bernhard(2012)]
{HB2012} Huemmerich, S., \& Bernhard, K.\ 2012, Open European Journal on Variable Stars, 149, 1 

\bibitem[Humphreys et al.(1972)]
{Hue1972} Humphreys, R.~M., Strecker, D.~W., \& Ney, E.~P.\ 1972, \apj, 172, 75

\bibitem[Jim{\'e}nez-Esteban et al.(2006)]
{JE2006} Jim{\'e}nez-Esteban, F.~M., Garc{\'\i}a-Lario, P., Engels, D., et al.\ 2006, \aap, 446, 773

\bibitem[Kleinmann et al.(1978)]
{Kle1978} Kleinmann, S.~G., Dickinson, D.~F., \& Sargent, D.~G.\ 1978, \aj, 83, 1206

\bibitem[Koralesky et al.(1998)]
{Koe1998} Koralesky, B., Frail, D.~A., Goss, W.~M., Claussen, M.~J., \& Green, A.~J.\ 1998, \aj, 116, 1323 

\bibitem[Kukarkin et al.(1971)]
{Kue1971} Kukarkin, B.~V., Kholopov, P.~N., Pskovsky, Y.~P., et al.\ 1971, General Catalogue of Variable Stars, 3rd ed.~(1971),  

\bibitem[Kwok et al.(1997)]
{Kwe1997} Kwok, S., Volk, K., \& Bidelman, W.~P.\ 1997, \apjs, 112, 55

\bibitem[Le Bertre, \& Nyman(1990)]
{LN1990} Le Bertre, T., \& Nyman, L.-A.\ 1990, \aap, 233, 477

\bibitem[Le Bertre et al.(2001)]
{Lee2001} Le Bertre, T., Matsuura, M., Winters, J.~M., et al.\ 2001, \aap, 376, 997 

\bibitem[Le Bertre et al.(2003)]
{Lee2003} Le Bertre, T., Tanaka, M., Yamamura, I., et al.\ 2003, \aap, 403, 943

\bibitem[Lewis et al.(1995)]
{Lee1995} Lewis, B.~M., David, P., \& Le Squeren, A.~M.\ 1995, \aaps, 111, 237 

\bibitem[Lloyd Evans, \& Catchpole(1989)]
{LC1989} Lloyd Evans, T., \& Catchpole, R.~M.\ 1989, \mnras, 237, 219

\bibitem[Lumsden et al.(2013)]
{Lue2013} Lumsden, S.~L., Hoare, M.~G., Urquhart, J.~S., et al.\ 2013, \apjs, 208, 11 

\bibitem[Matsunaga et al.(2005)]
{Mae2005} Matsunaga, N., Fukushi, H., \& Nakada, Y.\ 2005, \mnras, 364, 117 

\bibitem[Nguyen-Q-Rieu et al.(1979)]
{Nge1979} Nguyen-Q-Rieu, Laury-Micoulaut, C., Winnberg, A., \& Schultz, G.~V.\ 1979, \aap, 75, 351 

\bibitem[Nyman et al.(1993)]
{Nye1993} Nyman, L.-A., Hall, P.~J., \& Le Bertre, T.\ 1993, \aap, 280, 551

\bibitem[Ogbodo et al.(2019)]
{Oge2019} Ogbodo et al., 2019, submitted 

\bibitem[Paschenko et al.(1971)]
{Pae1971} Paschenko, M., Slysh, V., Strukov, I., et al.\ 1971, \aap, 11, 482

\bibitem[Payne et al.(1988)]
{Pae1988} Payne, H.~E., Phillips, J.~A., \& Terzian, Y.\ 1988, \apj, 326, 368

\bibitem[Peretto et al.(2007)]
{Pee2007} Peretto, N., Fuller, G., Zijlstra, A., et al.\ 2007, \aap, 473, 207

\bibitem[P{\'e}rez-S{\'a}nchez et al.(2017)]
{PS2017} P{\'e}rez-S{\'a}nchez, A.~F., Tafoya, D., Garc{\'\i}a L{\'o}pez, R., et al.\ 2017, \aap, 601, A68

\bibitem[Persson et al.(1976)]
{Pee1976} Persson, S.~E., Frogel, J.~A., \& Aaronson, M.\ 1976, \apj, 208, 753 

\bibitem[Preite-Martinez(1988)]
{PM1988} Preite-Martinez, A.\ 1988, \aaps, 76, 317 

\bibitem[Qiao et al.(2016a)]
{Qie2016a} Qiao, H.-H., Walsh, A.~J., G{\'o}mez, J.~F., et al.\ 2016a, \apj, 817, 37 

\bibitem[Qiao et al.(2016b)]
{Qie2016b} Qiao, H.-H., Walsh, A.~J., Green, J.~A., et al.\ 2016b, \apjs, 227, 26 (Paper I)

\bibitem[Qiao et al.(2018)]
{Qie2018} Qiao, H.-H., Walsh, A.~J., Breen, S.~L., et al.\ 2018, \apjs, 239, 15 (Paper II)

\bibitem[Ramos-Larios et al.(2009)]
{RL2009} Ramos-Larios, G., Guerrero, M.~A., Su{\'a}rez, O., et al.\ 2009, \aap, 501, 1207

\bibitem[Reid \& Moran(1981)]
{RM1981} Reid, M.~J., \& Moran, J.~M.\ 1981, \araa, 19, 231  

\bibitem[Reid(2002)]
{Rei2002} Reid, M.~J.\ 2002, Cosmic Masers: From Proto-Stars to Black Holes, 206, 506 

\bibitem[Robitaille et al.(2008)]
{Roe2008} Robitaille, T.~P., Meade, M.~R., Babler, B.~L., et al.\ 2008, \aj, 136, 2413 

\bibitem[Saito et al.(2012)]
{Sae2012} Saito, R.~K., Minniti, D., Angeloni, R., et al.\ 2012, The Astronomer's Telegram, 4426, 1

\bibitem[Samus et al.(2009)]
{Sae2009} Samus, N.~N., Kazarovets, E.~V., Durlevich, O.~V., Kireeva, N.~N., \& Pastukhova, E.~N.\ 2009, VizieR Online Data Catalog, 1, 

\bibitem[Samus' et al.(2017)]
{Sae2017} Samus', N.~N., Kazarovets, E.~V., Durlevich, O.~V., Kireeva, N.~N., \& Pastukhova, E.~N.\ 2017, Astronomy Reports, 61, 80 

\bibitem[Sevenster et al.(1997a)]
{Sea1997} Sevenster, M.~N., Chapman, J.~M., Habing, H.~J., Killeen, N.~E.~B., \& Lindqvist, M.\ 1997a, \aaps, 122, 79 

\bibitem[Sevenster et al.(1997b)]
{Seb1997} Sevenster, M.~N., Chapman, J.~M., Habing, H.~J., Killeen, N.~E.~B., \& Lindqvist, M.\ 1997b, \aaps, 124, 509 

\bibitem[Sevenster et al.(2001a)]
{Sea2001} Sevenster, M.~N., van Langevelde, H.~J., Moody, R.~A., et al.\ 2001a, \aap, 366, 481 

\bibitem[Sevenster \& Chapman(2001b)]
{Seb2001} Sevenster, M.~N., \& Chapman, J.~M.\ 2001b, \apjl, 546, L119 

\bibitem[Sevenster(2002)]
{Sev2002} Sevenster, M.~N.\ 2002, \aj, 123, 2788


\bibitem[Soszy{\'n}ski et al.(2013)]
{Soe2013} Soszy{\'n}ski, I., Udalski, A., Szyma{\'n}ski, M.~K., et al.\ 2013, \actaa, 63, 21 

\bibitem[Stroh et al.(2018)]
{Ste2018} Stroh, M.~C., Pihlstr{\"o}m, Y.~M., Sjouwerman, L.~O., et al.\ 2018, \apj, 862, 153 

\bibitem[Su{\'a}rez et al.(2006)]
{Sue2006} Su{\'a}rez, O., Garc{\'{\i}}a-Lario, P., Manchado, A., et al.\ 2006, \aap, 458, 173 

\bibitem[Szczerba et al.(2007)]
{Sze2007} Szczerba, R., Si{\'o}dmiak, N., Stasi{\'n}ska, G., \& Borkowski, J.\ 2007, \aap, 469, 799 

\bibitem[te Lintel Hekkert et al.(1989)]
{Lie1989} te Lintel Hekkert, P., Versteege-Hensel, H.~A., Habing, H.~J., \& Wiertz, M.\ 1989, \aaps, 78, 399 

\bibitem[te Lintel Hekkert(1991)]
{Lin1991} te Lintel Hekkert, P.\ 1991, \aap, 248, 209

\bibitem[te Lintel Hekkert et al.(1991)]
{Lie1991} te Lintel Hekkert, P., Caswell, J.~L., Habing, H.~J., et al.\ 1991, \aaps, 90, 327 

\bibitem[Titmarsh et al.(2014)]
{Tie2014} Titmarsh, A.~M., Ellingsen, S.~P., Breen, S.~L., Caswell, J.~L., \& Voronkov, M.~A.\ 2014, \mnras, 443, 2923

\bibitem[Titmarsh et al.(2016)]
{Tie2016} Titmarsh, A.~M., Ellingsen, S.~P., Breen, S.~L., Caswell, J.~L., \& Voronkov, M.~A.\ 2016, \mnras, 459, 157 

\bibitem[Turner(1969)]
{Tur1969} Turner, B.~E.\ 1969, \apj, 157, 103

\bibitem[Uscanga et al.(2012)]
{Use2012} Uscanga, L., G{\'o}mez, J.~F., Su{\'a}rez, O., \& Miranda, L.~F.\ 2012, \aap, 547, A40 

\bibitem[Vassiliadis et al.(1993)]
{Vassiliadis93} Vassiliadis, E. \& Wood, P.~R.\ 1993, \apj, 413, 641,

\bibitem[Walsh et al.(2009)]
{Wae2009} Walsh, A.~J., Breen, S.~L., Bains, I., \& Vlemmings, W.~H.~T.\ 2009, \mnras, 394, L70 

\bibitem[Walsh et al.(2011)]
{Wae2011} Walsh, A.~J., Breen, S.~L., Britton, T., Brooks, K. J., et al.\ 2011, \mnras, 416, 1764

\bibitem[Walsh et al.(2012)]
{Wae2012} Walsh, A.~J., Purcell, C., Longmore, S., Jordan, C.~H., \& Lowe, V.\ 2012, \pasa, 29, 262 

\bibitem[Walsh et al.(2014)]
{Wae2014} Walsh, A.~J., Purcell, C.~R., Longmore, S.~N., et al.\ 2014, \mnras, 442, 2240 

\bibitem[Weaver et al.(1965)]
{Wee1965} Weaver, H., Williams, D.~R.~W., Dieter, N.~H., \& Lum, W.~T.\ 1965, \nat, 208, 29 

\bibitem[Westerlund, \& Olander(1978)]
{WO1978} Westerlund, B.~E., \& Olander, N.\ 1978, \aaps, 32, 401

\bibitem[Wilson et al.(1972)]
{Wie1972} Wilson, W.~J., Schwartz, P.~R., Neugebauer, G., et al.\ 1972, \apj, 177, 523

\bibitem[Wilson et al.(2011)]
{Wie2011} Wilson, W.~E., Ferris, R.~H., Axtens, P., et al.\ 2011, \mnras, 416, 832 

\bibitem[Zijlstra et al.(1989)]
{Zie1989} Zijlstra, A.~A., te Lintel Hekkert, P., Pottasch, S.~R., et al.\ 1989, \aap, 217, 157 

\bibitem[Zijlstra et al.(2001)]
{Zie2001} Zijlstra, A.~A., Chapman, J.~M., te Lintel Hekkert, P., et al.\ 2001, \mnras, 322, 280 

\bibitem[Zoonematkermani et al.(1990)]
{Zoe1990} Zoonematkermani, S., Helfand, D.~J., Becker, R.~H., et al.\ 1990, \apjs, 74, 181

\end{thebibliography}
\end{document}